\documentclass[fullpaper,final]{nldl}

\paperID{34}

\vol{352}

\usepackage{mathtools}
\usepackage{bm}
\usepackage{graphicx}

\usepackage{booktabs}
\usepackage{threeparttable}
\usepackage{multirow}
\usepackage{enumitem}
\usepackage{makecell}

\usepackage{algorithm}
\usepackage{algorithmic}

\usepackage{eurosym}

\usepackage{listings}
\usepackage{hyperref}
\usepackage{url}
\hypersetup{
  pdfusetitle,
  colorlinks,
  linkcolor = BrickRed,
  citecolor = NavyBlue,
  urlcolor  = Magenta!80!black,
}

\title{A Station-Based Evaluation of Machine Learning-based Weather Forecasting Models in Northern Norway
}
\author[1]{Siyan Chen}
\author[1]{Lars Uebbing}
\author[1,5]{Eirik Mikal Samuelsen}
\author[1]{Georgios Leontidis}
\author[2]{Arnt-Børre Salberg}
\author[1,3]{Sébastien Lefèvre}
\author[1,2,4]{Robert Jenssen}
\author[1]{Kristoffer Wickstrøm}
\affil[1]{UiT The Arctic University of Norway}
\affil[2]{Norwegian Computing Center}
\affil[3]{University of South Brittany}
\affil[4]{University of Copenhagen}
\affil[5]{The Norwegian Meteorological Institute}

\begin{document}
\maketitle

\begin{abstract}
Recent machine learning weather prediction (MLWP) models have demonstrated remarkable forecasting skill on global reanalysis-based benchmarks. However, their performance remains unclear in challenging environments such as Northern Norway, where narrow fjords and rapidly changing weather result in highly variable local wind conditions. In this case study, we evaluate FourCastNet3 (FCN3), GraphCast, and ECMWF High Resolution Forecast (HRES) for wind speed forecasting using multi-year station observations from Northern Norway, focusing on their relative performance, generalization beyond the training period, and performance under high-wind conditions. Our results show that HRES slightly outperforms FCN3 and GraphCast, with an overall RMSE of 2.89 m\,s$^{-1}$, compared to 2.96 m\,s$^{-1}$ for FCN3 and 2.94 m\,s$^{-1}$ for GraphCast. Notably, the MLWP models maintain comparable performance beyond their respective training periods, with no clear evidence of noticeable degradation over the evaluated years. FCN3 performs best under high-wind conditions, although all models substantially underestimate strong winds. Our findings suggest that MLWP has become competitive with NWP for local wind, but further refinements are still needed to capture complex terrain better.   
\end{abstract}

\section{Introduction}\label{sec:introduction}
Accurate wind forecasting plays a critical role in a wide range of real-world applications, including transportation~\citep{liu2024windtrans, alves2023automated}, renewable energy production~\citep{foley2012current, tuncar2024review}, maritime operations~\citep{kytariolou2022ship, moon2024post}, and disaster risk management~\citep{camps2025artificial, jain2023leveraging}. Current forecasting methods mainly include Numerical Weather Prediction (NWP) and Machine Learning-based Weather Prediction (MLWP) approaches ~\citep{waqas2025artificial}. Although NWP models such as the Integrated Forecasting System (IFS) developed by the European Centre for Medium-Range Weather Forecasts (ECMWF)~\citep{owens2018ecmwf} and the Global Forecast System (GFS) provide skillful forecasts and remain the foundation of operational forecasting, they require prohibitive computational costs and are slow to produce forecasts~\citep{owens2018ecmwf, fang2026efficientparametercalibrationnumerical, ben2024rise}. For example, the IFS operates on 293 million grid points and assimilates 40 million observations daily, taking 2 hours and 10 minutes to generate a 10-day high-resolution global forecast~\citep{wedi2015supercomputing}. 
\begin{figure}[tb]
  \centering
  \includegraphics[width=\linewidth]{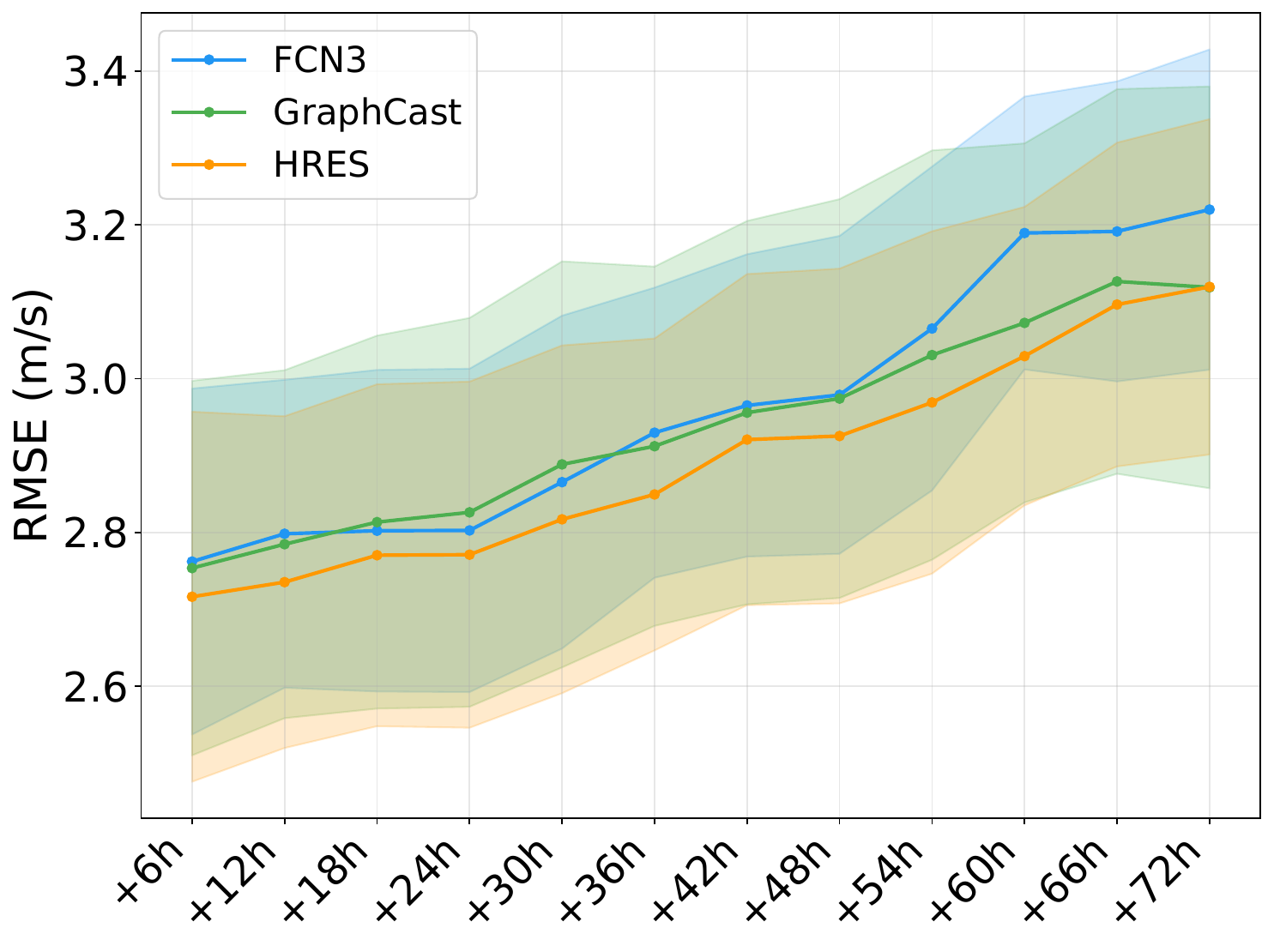}
  \caption{Mean RMSE of wind speed forecasts from FCN3, GraphCast, and HRES at lead times from +6 to +72 hours with shaded bands indicating $\pm1$ standard error, calculated using station observations as the reference. Results are aggregated across all 12 stations and all evaluation years from 2016 to 2022.}
  \label{fig:forecast-rmse-lead-time-obs}
  \vspace{-15pt}
\end{figure}
With the rapid advancement of artificial intelligence (AI), recent studies have shown that state-of-the-art (SOTA) MLWP models have achieved comparable or even superior performance in terms of accuracy, while significantly improving computational efficiency and costs~\citep{ben2024rise, nguyen2025atmosarena}. These models provide a data-driven alternative to conventional NWP without numerically solving the governing physical equations and are typically trained on decades of gridded reanalysis datasets such as ERA5, the fifth-generation atmospheric reanalysis produced by ECMWF~\citep{hersbach2020era5}, to learn the mapping from current atmospheric conditions to future states~\citep{waqas2025artificial, zhang202570}. Representative models include FourCastNet 3 (FCN3)~\citep{Bonev2025FourCastNet3A}, GraphCast~\citep{lam2023learning}, and Pangu-Weather~\citep{bi2023accurate}, among others~\citep{kurth2023fourcastnet, bonev2023spherical, chen2023fuxi, chen2025operational, pmlr-v202-nguyen23a, price2025probabilistic, chantry2025aifs}. However, it remains unclear how well MLWP models perform for complex terrain at station levels where high resolution is necessary, as most existing evaluations are conducted on global gridded benchmark datasets and few studies test against in situ measurements, leaving the practical implications of these MLWP models for local prediction insufficiently understood~\citep{rasp2020weatherbench, rasp2024weatherbench}. Moreover, another limitation of these global weather models is their inability to resolve fine-scale topography and the resulting local weather variations, particularly wind forecasts~\citep{ImprovingtheRepresentation, martinez2021review, lin2026reconstructing, haji2025beyond}. This limitation is not unique to MLWP models, but also affects conventional NWP global forecasting systems. For example, \citet{haiden2018addressing} identified persistent biases in ECMWF’s medium-range forecasts of near-surface weather parameters, such as 2-meter temperature and 10-meter wind speed, and such biases are associated with the coupling between the atmosphere and the land surface in the IFS. This challenge has also been demonstrated in Northern Norway, where surface wind simulated by the Weather Research and Forecasting (WRF) model was found to be strongly affected by the choice of atmospheric input data and horizontal grid spacing in complex terrain~\citep{solbakken2021evaluation}. Northern Norway, characterized by mountainous coastlines, narrow fjords, and rapidly changing weather~\citep{jonassen2012multi}, provides a natural and challenging testbed for this purpose. Furthermore, recent evidence suggests that MLWP models may remain anchored to the climate represented in their historical training data, resulting in systematic biases when applied to more recent atmospheric states~\citep{landsberg2026forecasting}. To better understand these limitations, our study is guided by the following research questions:
\begin{itemize}
    \item \textbf{RQ1:} How accurately do machine learning-based and physics-based models predict 10 m wind speed in Northern Norway?
    \item \textbf{RQ2:} How does forecast performance vary across training and non-training years?
    \item \textbf{RQ3:} To what extent is forecast performance affected by lead time, station location, and high-wind events?
\end{itemize}
These research questions are practically critical because the value of a weather forecast ultimately depends on its reliability at specific locations~\citep{yang2025local}, decision-relevant lead times~\citep{Tim2019verification}, and during high-wind situations~\citep{price2025probabilistic}, as well as on whether its performance remains robust beyond the training period despite possible temporal distribution shifts associated with a changing climate~\citep{zhang2026physics, landsberg2026forecasting}. 

We selected FCN3 and GraphCast based on their strong forecasting performance, architectural diversity, and publicly available pretrained weights~\citep{rasp2024weatherbench, Bonev2025FourCastNet3A, lam2023learning}. FCN3 employs spherical neural operators~\citep{Bonev2025FourCastNet3A}, whereas GraphCast uses a graph neural network~\citep{lam2023learning}. Transformer-based models, including Pangu-Weather and Aurora, are also considered, but GraphCast was preferred over Pangu-Weather according to its reported lower RMSE for 99.2\% of 252 evaluated targets~\citep{lam2023learning}, and Aurora's heterogeneous pretraining data would introduce additional differences~\citep{bodnar2025foundation}. Choosing both models trained and initialized on ERA5 allows us to assess whether their strong performance on gridded global benchmarks also extends to station-level wind forecasting, rather than an exhaustive ranking of all MLWP models. We compare their forecasts with HRES, which serves as a well-established physics-based reference and a natural numerical forecasting baseline. An overview of model performance across forecast lead times is presented in Fig.~\ref{fig:forecast-rmse-lead-time-obs}. 
Rather than proposing a new MLWP method, our study focuses on systematically evaluating global leading MLWP models against station observations in complex terrain, aiming to identify strengths and weaknesses that can inform future developments in the broader MLWP community.
Our contributions and key findings of this work are:
\begin{enumerate}
    \item A comprehensive station-level evaluation of two SOTA MLWP models and HRES on 12 weather stations in northern Norway.
    \item We characterize the relative performance of HRES, FCN3, and GraphCast across forecast lead times, stations, and wind force categories, with particular emphasis on high-wind conditions. 
    \item Extensive experimental results show that overall, HRES achieves the lowest root mean square error (RMSE), while FCN3 exhibits the smallest bias and performs best under high-wind conditions. All models substantially underestimate high wind speeds.
\end{enumerate}

\section{Related Work}
\paragraph{Traditional weather forecasting.} Unlike MLWP models learning and predicting weather patterns from large-scale historical datasets, NWP combines physical equations, numerical solvers, observations, and data assimilation to predict the future atmospheric state~\citep{bauer2015quiet}. Operational global NWP systems include IFS~\citep{owens2018ecmwf}, GFS~\citep{noaaGFS}, and several other forecasting systems~\citep{zangl2015icon, rogers2009ncep, cote1998operational, benjamin2016north}. These systems mainly differ in their dynamical cores and grid structures, with IFS using a spectral-transform core on a reduced Gaussian grid~\citep{owens2018ecmwf} and GFS using a finite-volume code on a cubed-sphere grid~\citep{noaaGFS}. At regional scales, models such as the Weather Research and Forecasting (WRF) model~\citep{powers2017weather} and HARMONIE--AROME~\citep{bengtsson2017harmonie} provide higher-resolution forecasts intended to represent mesoscale and terrain-related processes. However, increased horizontal resolution does not fully eliminate the representativeness gap between gridded model output and station observations~\citep{koltzow2019nwp}. Unresolved geography and topography and station-to-grid mismatch remain important sources of near-surface wind errors in complex terrain, even in high-resolution NWP models~\citep{ImprovingtheRepresentation,jimenez2013ability,zhang2011improved,solbakken2021evaluation,valkonen2020evaluation}.
\vspace{-0.25cm}
\paragraph{Machine learning-based weather forecasting models.}
Early data-driven weather forecasting primarily relied on statistical time-series models and conventional machine learning methods~\citep{shi2025deep,haji2025beyond}. The availability of large-scale reanalysis datasets and advances in deep learning subsequently enabled global MLWP models with diverse architectures. FourCastNet introduces the Adaptive Fourier Neural Operator at $0.25^\circ$ resolution~\citep{kurth2023fourcastnet}, and FCN2 further applies a spherical variant to better represent Earth's geometry~\citep{pmlr-v202-bonev23a}. Pangu-Weather uses a 3D Earth-specific transformer~\citep{bi2023accurate}, Graphcast a multi-mesh graph neural network~\citep{lam2023learning}, FuXi a cascaded architecture for 15-day forecasts~\citep{chen2023fuxi}, and ECMWF's AIFS a graph-based encoder--decoder with a transformer processor~\citep{chantry2025aifs}. 

More recent work has extended towards probabilistic predictions~\citep{price2025probabilistic, Bonev2025FourCastNet3A, atmos16010082}, such as U-Cast~\citep{ruhling2026u}, OmniCast~\citep{nguyen2026omnicast} and ATLAS~\citep{kossaifi2026demystifying}. GenCast uses conditional diffusion to generate probabilistic ensemble forecasts~\citep{price2025probabilistic}, while FCN3 employs a geometric convolutional architecture for scalable probabilistic forecasting on the sphere~\citep{Bonev2025FourCastNet3A}. Aurora follows a foundation model approach, using pre-training on diverse Earth-system datasets before adaptation to specific forecasting tasks~\citep{bodnar2025foundation}. Similarly, Earth Physics Transformer 1.5 (EPT-1.5) and its successor, EPT-2, are Transformer-based Earth-system foundation models developed for global forecasting, with an emphasis on energy-relevant variables such as near-surface wind, temperature, and solar radiation~\citep{molinaro2024ept15technicalreport, molinaro2025ept2technicalreport}. In contrast, Aardvark learns an end-to-end mapping from observations to global gridded and local station forecasts, reducing its dependence on NWP-generated initial states~\citep{allen2025end}. These models show a progression from deterministic forecasting toward probabilistic, transferable, and end-to-end approaches for weather prediction and Earth observation and forecasting systems~\citep{shi2025deep, allen2025end, allen2026machine}. While most data-driven weather forecasting models have been developed at the global scale, recent studies have increasingly explored regional approaches to provide higher-resolution predictions~\citep{wijnands2026data}. These include graph-based limited-area models~\citep{oskarsson2023graph,adamov2025building}, stretched-grid approaches~\citep{nordhagen2025high,nipen2026regional}, with YingLong~\citep{xu2025artificial} explicitly integrating topographic information and ~\citet{nipen2026regional} highlighting the importance of higher spatial resolution for forecasting over complex terrain and coastlines. 

More generally, time-series forecasting applies diverse methods to capture temporal and multivariate dependencies. PatchTST~\citep{nie2022time} uses patch-based representations, while MoFo~\citep{ma2026mofo}, PHAT~\citep{ma2026phat}, and TimesNet~\citep{wu2022timesnet} model periodic patterns. Crisp~\citep{wangcrisp} captures frequency-adaptive multivariate interactions, whereas TimeMixer~\citep{wang2024timemixer} models seasonal and trend patterns across temporal resolutions.
\vspace{-0.25cm}
\paragraph{Evaluation of global MLWP models.}
Most AI weather models are evaluated globally against ERA5, with HRES commonly used as the main physics-based baseline~\citep{rasp2020weatherbench, rasp2024weatherbench}. WeatherBench 2 further standardizes this evaluation against ERA5 reanalysis by using grid-based metrics across variables and forecast lead times~\citep{rasp2024weatherbench}. Such studies have shown that MLWP models can achieve competitive global forecast skill, but have also revealed limitations in temporal generalization and extreme-event prediction~\citep{landsberg2026forecasting, pasche2025validating, zhang2026physics}. For example, \citet{landsberg2026forecasting} found that FourCastNet and Pangu-Weather produced systematic cold biases in 2020--2025 boreal-winter temperature forecasts, with their predictions resembling climates from approximately 15--20 years earlier, and others also reported degraded performance during high-impact extreme events~\citep{pasche2025validating,zhang2026physics}.
In addition, gridded evaluation against ERA5 cannot fully determine whether this global performance translates to out-of-grid point locations, particularly because ERA5 itself represents a model-based gridded reconstruction of the atmosphere rather than independent point observations and is widely used for training MLWP systems as well~\citep{hersbach2020era5,rasp2024weatherbench,jin2024weatherrealbenchmarkbasedinsitu}. 

To provide a more direct assessment of near-surface forecast performance, several recent studies have compared model outputs with station observations~\citep{ramavajjala2023verification, bodnar2025foundation,molinaro2025ept2technicalreport, ben2024rise, bouallegue2026realistic}. Aurora and EPT-2 were evaluated for 10-m wind speed and 2-m temperature using observations from more than 13,000 and 14,000 stations worldwide, respectively~\citep{bodnar2025foundation,molinaro2025ept2technicalreport}. Similarly, Pangu-Weather was evaluated for the same variables at 183 Norwegian surface synoptic stations for lead times of up to 60 hours~\citep{bremnes2024evaluation}. Nevertheless, station-level evaluations remain limited in regions with highly complex terrain, where unresolved mountains, fjords, and coastlines can strongly influence local wind conditions~\citep{solbakken2021evaluation,koltzow2019nwp}. Northern Norway therefore provides a natural testbed for examining whether the global skill of MLWP models translates into reliable station-level wind forecasts under such challenging conditions.
\begin{figure*}[tb]
    \centering
    \includegraphics[width=\linewidth]{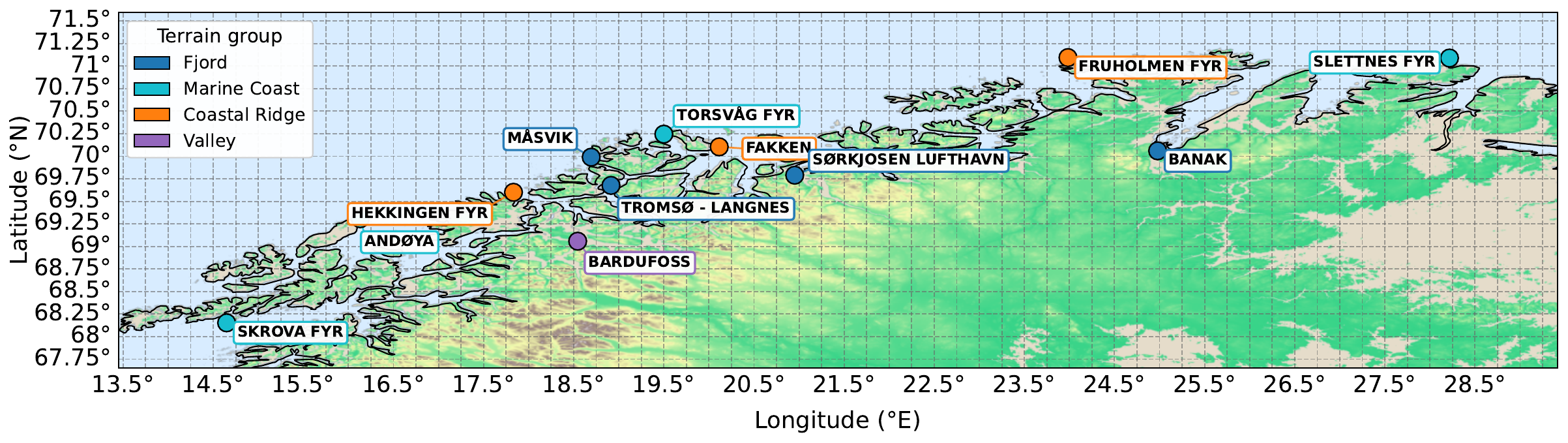}
    \caption{Locations of the 12 meteorological stations in Northern Norway, grouped by terrain category. Grid lines indicate \(0.25^\circ \times 0.25^\circ\) cells.}
    \label{fig:station_map}
\end{figure*}

\section{Station-Based Evaluation of Weather Forecasting Models}
This section presents the framework used to evaluate global weather forecasts at local observation sites. Building on the research questions introduced in Section~\ref{sec:introduction}, our evaluation considers overall accuracy, temporal robustness, and performance variations across lead times, station locations, and high-wind events. 
\subsection{Evaluation Framework}
We apply a station-based framework to assess FCN3 and GraphCast. The forecasts of these two models are generated using an open-source NVIDIA Earth2Studio framework that provides a unified workflow for running several AI-based weather and climate models~\citep{earth2studio}, whereas the $0.25^\circ$ regridded HRES forecast data were obtained from WeatherBench 2~\citep{rasp2024weatherbench}. All forecast and reference datasets are then mapped to a common set of station locations and valid times using the procedure described in the following. Within this benchmark, we compare the three forecast products against in-situ station observations to assess their local forecast performance. 
\vspace{-0.25cm}
\paragraph{Wind speed calculation.}
For FCN3, GraphCast, and ERA5, 10-m wind speed is calculated from the zonal and meridional components as $w = \sqrt{u_{10}^{2}+v_{10}^{2}}$, where $u_{10}$ and $v_{10}$ are the eastward and northward wind components at 10~m above the surface. For HRES, we use the precomputed 10-m wind-speed variable provided by WeatherBench 2.
\paragraph{Spatial and temporal matching.}
Following the in-situ benchmark approach implemented by Microsoft~\citep{jin2024weatherrealbenchmarkbasedinsitu}, the gridded 10-m wind speed outputs from the three selected models are interpolated to the coordinates of each station. Because all products are defined on regular latitude--longitude grids, this is performed using bilinear interpolation from the four surrounding grid points. The same procedure is applied consistently to all gridded products, yielding a wind speed time series for each model and reference dataset at every station location. Given an initialization input at time \(t_{\mathrm{init}}\) and a specified lead time \(t_{\mathrm{lead}}\), the corresponding valid time is defined as $t_{\mathrm{valid}} = t_{\mathrm{init}} + t_{\mathrm{lead}}$. The interpolated forecast values are matched with station observations at the exact valid time. The intersection of the resulting station--time pairs is then retained to ensure that FCN3, GraphCast, and HRES are evaluated using the same samples at each forecast lead. Forecast errors are then calculated separately against station observations by using $ e_i^{\mathrm{obs}} = \hat{w}_i-w_i^{\mathrm{obs}}$, where $\hat{w}_i$ denotes the forecast wind speed and $w_i^{\mathrm{obs}}$ the corresponding station observation. The resulting metrics therefore quantify model performance against in situ measurements.
\vspace{-0.25cm}
\paragraph{Forecast generation.}
FCN3 and GraphCast forecasts are initialized from the ERA5 reanalysis dataset at a spatial resolution of $0.25^\circ$, corresponding to a global grid of $1,440 \times 721$ latitude-longitude points. For each initialization, forecasts are produced via a single multi-step autoregressive rollout, producing the complete forecast trajectory defined by the number of lead times. Notably, GraphCast is a deterministic forecasting model,
whereas FCN3 is probabilistic. For FCN3, we evaluate a single stochastic sample with a fixed default random seed of 333, rather than an ensemble mean. To keep processing and storage tractable across the seven-year evaluation period, forecasts are generated in weekly batches, and the raw global output from each run is immediately bilinearly interpolated to the twelve station locations. The full global fields are then discarded, while the station-level forecasts are stored as weekly Zarr archives with dimensions corresponding to station, initialization time, and lead time. 
\vspace{-0.1cm}
\paragraph{Initialization.} 
All forecasts are initialized at 00:00~UTC and evaluated every 6~h from $+6$ to $+72$~h, with the lead times $\tau \in \{6,12,18,\ldots,72\}\ \mathrm{h}$. This produces 12 forecast lead times for each forecast cycle. Forecasts are matched to the reference datasets according to their valid time. For example, a forecast initialized at 00:00~UTC with a lead time of $+12$~h is compared with station observations at 12:00~UTC on the same day.
\vspace{-0.1cm}
\paragraph{High-wind evaluation.}
Extreme winds in Europe caused severe damage worth \euro 138 billion over the period 1981–2016, and Norway is no exception, with windstorms accounting for more than half of the insurance payouts from the Norwegian Natural Perils Pool~\citep{jaison2025towards}. Given these societal and economic impacts, model performance under high-wind conditions is evaluated separately. A high-wind case is defined when the observed wind speed at the station satisfies $ w_i^{\mathrm{obs}} \geq 10.8~\mathrm{m\,s^{-1}}$, using the threshold adopted by a Norwegian online weather forecasting service called Yr for high winds~\citep{yr_high_winds}. The evaluation metrics are recalculated for this high-wind subset. Scatter plots of forecast and observed wind speeds, together with the \(1{:}1\) reference line, are used to assess agreement across the wind speed range and identify systematic overestimation or underestimation under high-wind conditions.
\subsection{Experimental Setup}
We evaluate three global forecast products using in situ observations from 12 stations in Northern Norway during 2016--2022. The stations were selected to provide broad geographical coverage, sufficient observations, and good exposure, minimizing the influence of nearby obstacles such as tall trees on 10-m wind measurements. The study period was determined by the availability of the HRES archive provided by WeatherBench 2~\citep{rasp2024weatherbench}. Forecasts and reference data are spatially and temporally aligned before forecast errors are calculated using a common set of metrics. Code is available at \url{https://github.com/chensy618/WF-experiments}. 

\paragraph{Study area.}
We consider 12 meteorological stations distributed across Northern Norway, covering approximately $68.15$--$71.09^{\circ}$N and $14.65$--$28.22^{\circ}$E. The stations represent locations with different coastal exposure and surrounding terrain. Figure~\ref{fig:station_map} shows their geographical distribution, while zoomed-in views of the regional topography are provided in Fig.~\ref{fig:station_zoom_in} in Appendix~\ref{sec:weather_station_info}.
\paragraph{Evaluation metrics.}
Forecast performance is quantified using RMSE and bias, two commonly used metrics in weather forecast verification that evaluate the overall magnitude of forecast errors and systematic forecast bias~\citep{cawcr_verification,wilks2011statistical,rasp2024weatherbench}, respectively.
\[
\mathrm{RMSE} = \sqrt{\frac{1}{N}\sum_{i=1}^{N}\left(e_i^{\mathrm{obs}}\right)^2},
\qquad
\mathrm{Bias} = \frac{1}{N}\sum_{i=1}^{N}e_i^{\mathrm{obs}},
\]
\noindent where $N$ is the number of valid forecast-reference pairs. Positive bias indicates systematic overestimation, while negative bias indicates underestimation. Normalized RMSE (nRMSE) and normalized bias (nBias) are additionally reported to account for differences in the characteristic wind-speed magnitude among stations and to facilitate relative comparisons of forecast errors across locations~\citep{drechsel2012wind}. They are calculated relative to the mean wind speed $\bar{w}$ of the corresponding reference within each evaluation subset:
\[
\mathrm{nRMSE} = \frac{\mathrm{RMSE}}{\bar{w}}\times 100\%,
\qquad
\mathrm{nBias} = \frac{\mathrm{Bias}}{\bar{w}}\times 100\%.
\]
We also report the Pearson correlation coefficient between observed and predicted wind speeds to assess their overall agreement.
The metrics are first calculated across all stations and years to summarize overall model performance. They are then computed separately for each forecast lead time, station, and year to examine how forecast skill varies across temporal horizons, locations, and evaluation periods. Moreover, the anomaly correlation coefficient (ACC) is used to assess how well forecasts capture deviations from climatology:
\begin{equation}
    \mathrm{ACC} = \mathrm{Corr}(F-C,\ O-C),
\end{equation}
where $F$, $O$, and $C$ denote forecast values, reference values,
and the climatological mean, respectively.
For station-based evaluation, we calculate station-specific
day-of-year climatologies from 2023--2025 observations using
a 61-day sliding window, following WeatherBench 2~\citep{rasp2024weatherbench}. To further investigate how forecast performance varies with wind speed categories, we complement the continuous error metrics with categorical verification based on the Beaufort scale~\citep{wmo_beaufort,infoplaza_beaufort}. It defines 13 wind-force classes, ranging from 0 to 12, and corresponding wind-speed ranges are provided in Appendix~\ref{sec:beaufort-force}. We apply the Equitable Threat Score (ETS)~\citep{cawcr_verification} as a categorical measure of forecast skill that corrects for the number of hits expected by random chance. For each wind-speed category $k$, the ETS is defined as:
\begin{equation}
\mathrm{ETS}_k =
\frac{H_k-H_{k,\mathrm{random}}}
{H_k+M_k+F_k-H_{k,\mathrm{random}}},
\end{equation}
where the expected random hits are:
\begin{equation}
H_{k,\mathrm{random}}=\frac{(H_k+M_k)(H_k+F_k)}{N}.
\end{equation}
Here, $H_k$, $M_k$, and $F_k$ denote the numbers of hits, misses, and false alarms for category $k$, respectively, and $N$ is the total number of forecast--observation pairs. ETS ranges from negative values to 1, with higher values indicating better categorical forecast skill.

\begin{table*}[tb]
\centering
\caption{Model performance evaluated against station observations across all stations during 2016--2022.}
\label{tab:overall_metrics}
\setlength{\tabcolsep}{6pt}
\begin{threeparttable}
\begin{tabular}{clccccc}
\multicolumn{7}{r}{$N=366{,}585$} \\
\toprule
Model
& RMSE $\downarrow$
& nRMSE $\downarrow$
& Bias $(\to 0)$
& nBias $(\to 0)$ & Corr. & ACC\\
\midrule

HRES
& $\bm{2.89\,(0.75)}$
& $\bm{53.06\,(15.00)}$
& $-1.21\,(0.85)$
& $-21.29\,(15.76)$ & $\bm{.685(.088)}$ &  $\bm{.5654(.171)}$ \\

FCN3
& $2.96\,(0.71)$
& $54.27\,(14.20)$
& $\bm{-0.85\,(1.04)}$
& $\bm{-15.63\,(19.32)}$ & $.648(.088)$ & $.5504(.164)$\\

GC
& $2.94\,(0.86)$
& $53.43\,(14.61)$
& $-1.26\,(0.99)$
& $-21.40\,(16.68)$ & $.682(.095)$ & $.5651(.173)$ \\
\bottomrule
\end{tabular}
*\textbf{Bold} indicates the best result for each reference, and values are reported as the mean with the standard deviation given in parentheses. RMSE and Bias are in \(\mathrm{m\,s^{-1}}\), while nRMSE and nBias are in \%. GC: GraphCast.
\end{threeparttable}
\end{table*}
\begin{figure*}[tb]
    \centering
    \begin{subfigure}[t]{0.48\textwidth}
        \centering
        \includegraphics[width=\linewidth]{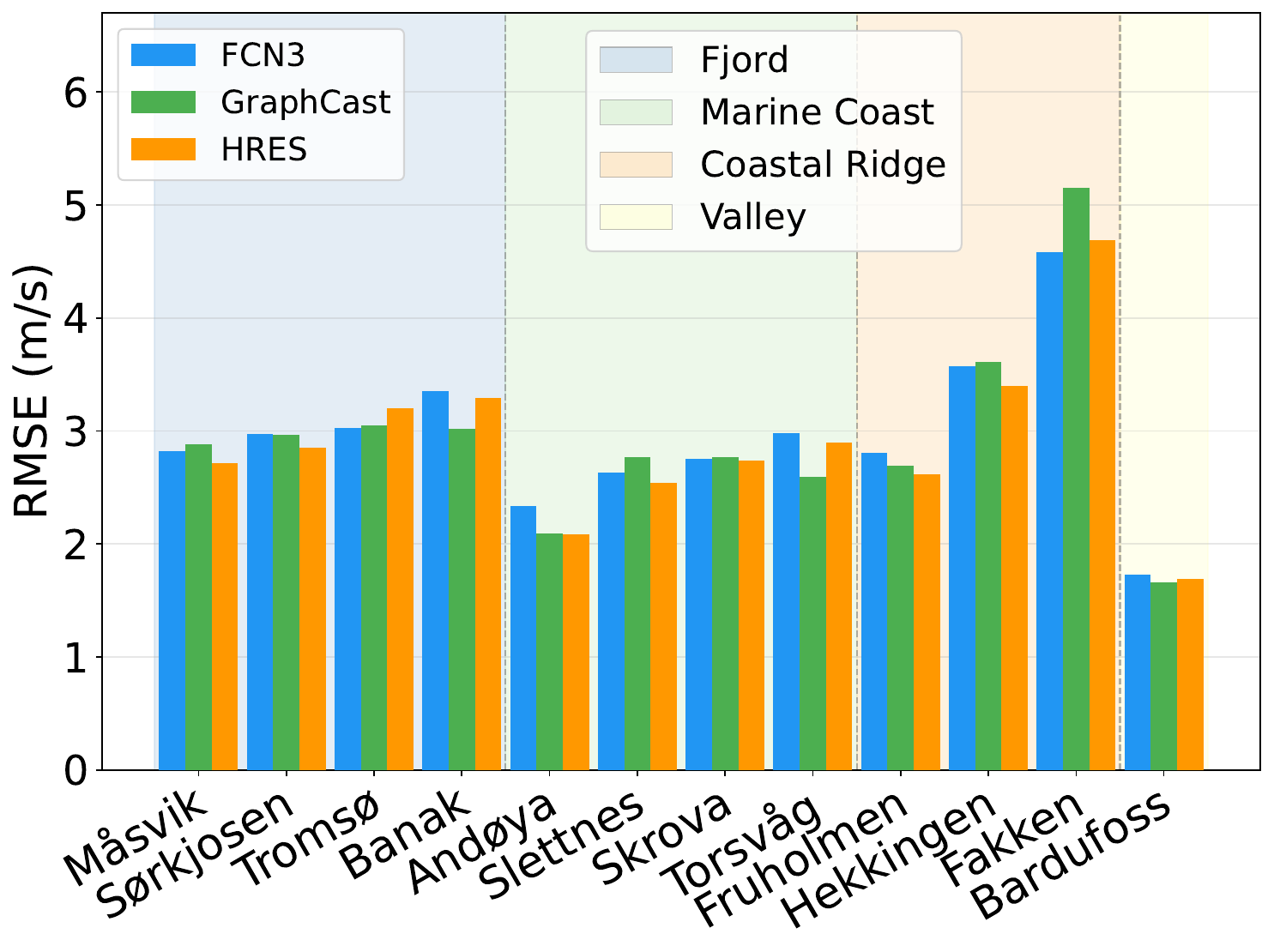}
        \caption{RMSE comparison across stations, grouped by terrain type as indicated by the background colors.}
        \label{fig:panel_rmse_station}
    \end{subfigure}
    \hfill
    \begin{subfigure}[t]{0.48\textwidth}
        \centering
        \includegraphics[width=\linewidth]{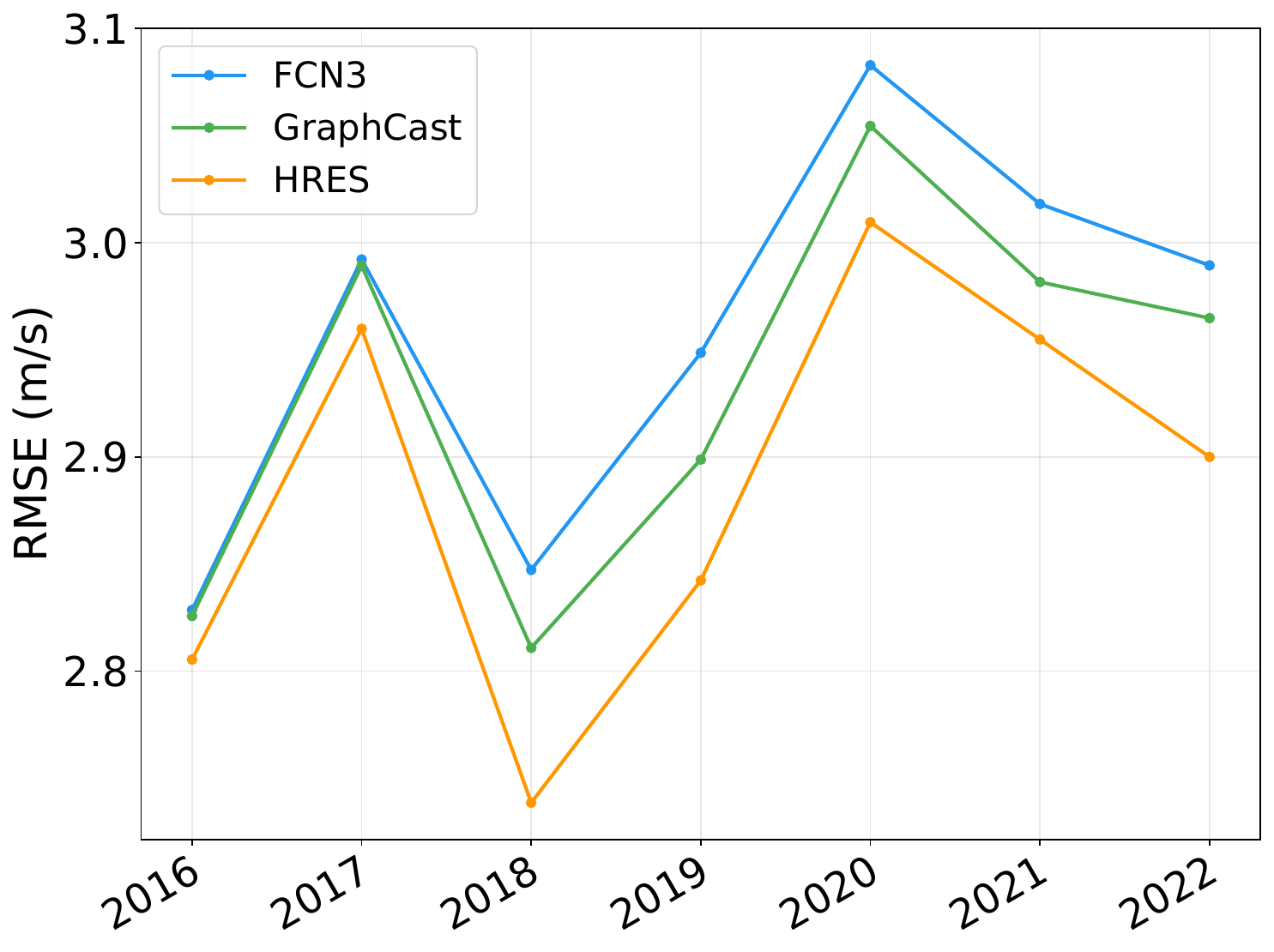}
        \caption{Annual mean RMSE for each model from 2016 to 2022.}
        \label{fig:panel_rmse_year}
    \end{subfigure}

    \caption{RMSE performance of FCN3, GraphCast, and HRES across forecast lead times, stations, and years.}
    \label{fig:three_panels}
\end{figure*}
\begin{table}[tb]
\centering
\caption{Pairwise station-level RMSE differences from 10,000 bootstrap resamples (GC: GraphCast).}
\label{tab:rmse_bootstrap}
\begin{tabular}{lccc}
\toprule
Models & Diff. & 95\% CI & Sig. \\
\midrule
FCN3--GC   & 0.029 & [-0.113, 0.157] & No \\
FCN3--HRES & 0.073 & \textbf{[0.005, 0.135]}  & Yes \\
GC--HRES   & 0.044 & [-0.074, 0.164] & No \\
\bottomrule
\end{tabular}
\end{table}
\section{Results}
In this section, we first present the overall results, followed by analyses across lead times, stations, evaluation years, and high-wind conditions.

\subsection{Overall Forecasting Performance}
Table~\ref{tab:overall_metrics} shows the model performance averaged over all stations from +6 h to +72 h during 2016-2022, where HRES achieved the lowest overall RMSE of 2.89 m\,s$^{-1}$ and nRMSE of 53.06\%, followed closely by GraphCast and FCN3 with RMSE values of 2.94 and 2.96 m\,s$^{-1}$, respectively. HRES also achieved the highest correlation and ACC (r=0.685, ACC=0.5654), whereas GraphCast (r=0.682, ACC= 0.5651) and FCN3 (r=0.648, ACC=0.5504) showed lower performance. The small differences indicate broadly comparable overall forecast accuracy among them, with HRES showing only a modest advantage. We further performed a station-level cluster bootstrap to assess the significance of the small differences. The bootstrap results in Table~\ref{tab:rmse_bootstrap} showed that only the FCN3--HRES difference was statistically significant. FCN3 had a \(0.07~\mathrm{m\,s^{-1}}\) higher RMSE than HRES (95\% CI: \(0.005\)--\(0.135~\mathrm{m\,s^{-1}}\)), while the confidence intervals for FCN3--GraphCast and GraphCast--HRES included zero. Thus, HRES showed only a small but statistically significant advantage over FCN3, and the other pairwise differences were not statistically significant. We also conducted additional moving block bootstrap~\citep{lahiri2013resampling, buhlmann2002bootstraps, hongyi1996bootstrapping} to account for temporal correlation and shared weather-event dependence across stations, and detailed results are provided in Appendix~\ref{app:block_bootstrap}. 

\begin{figure}[tb]
    \centering
    \includegraphics[width=\linewidth]{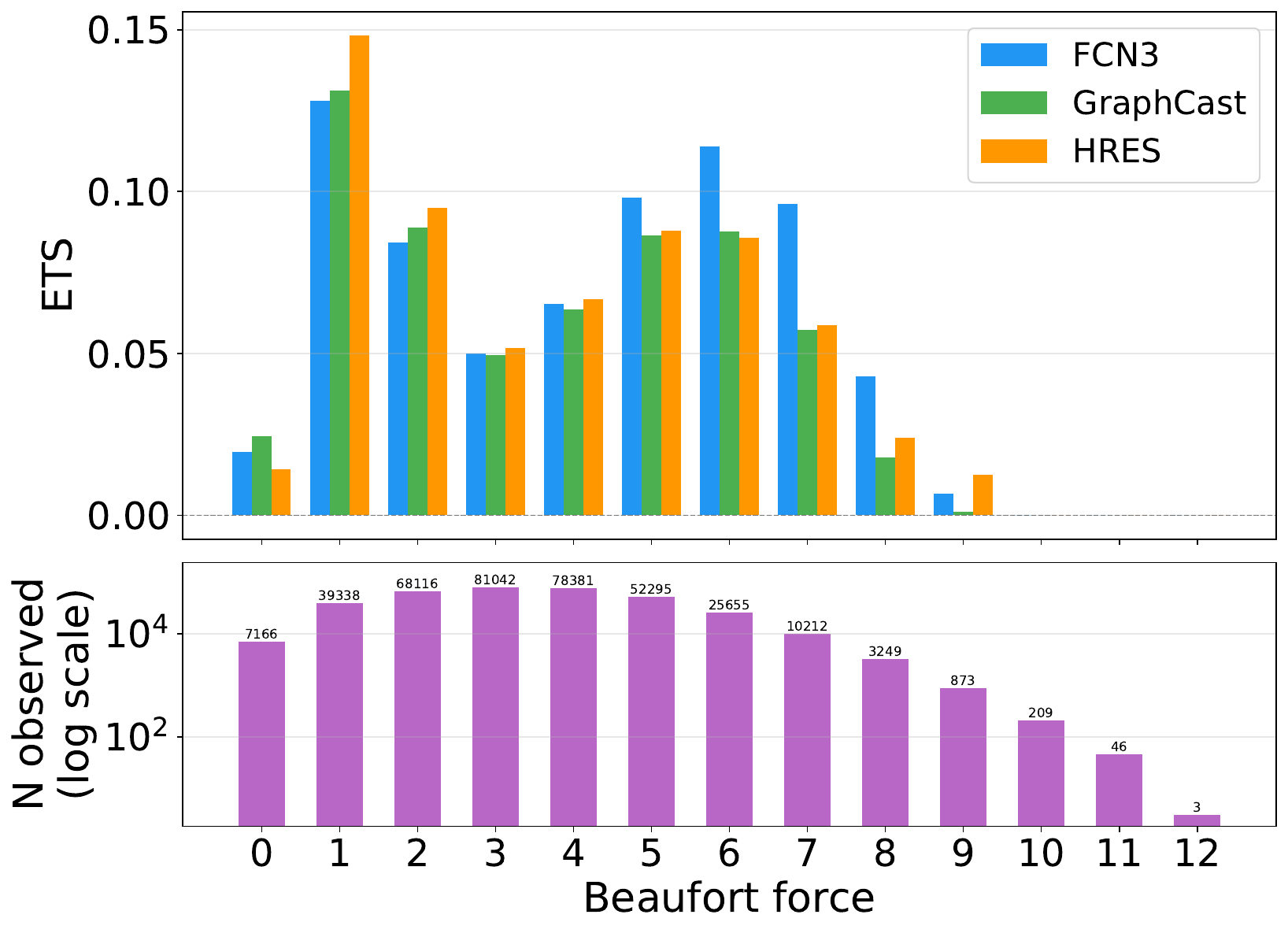}
    \caption{Model Performance on ETS by Beaufort force category}
    \label{fig:ETS}
\end{figure}

All models exhibited negative bias, indicating systematic underestimation of observed wind speeds. FCN3 showed the smallest bias at -0.85 m\,s$^{-1}$ and the lowest absolute nBias of 15.63\%, whereas HRES and Graphcast had larger negative biases of -1.21 m\,s$^{-1}$ and -1.26 m\,s$^{-1}$, respectively. Although HRES produced the lowest aggregate error, FCN3 better reproduced the overall magnitude of the observed wind speeds. This overall ranking is further examined across forecast lead times. In Fig.~\ref{fig:forecast-rmse-lead-time-obs}, HRES achieves the lowest RMSE from +6 h onward and RMSE increases with lead time for all models, with larger gaps at longer horizons. FCN3 generally ranks second up to +30 h, while GraphCast shows the largest errors over this range.  
ETS varies across Beaufort categories (Fig.~\ref{fig:ETS}). HRES performs best for forces 1--4, while FCN3 leads for forces 5--8. Skill drops sharply above force 8 and is negligible for forces 10--12. The lower panel validates this pattern by showing that observations are concentrated in the lower Beaufort categories, while stronger-wind events are much less frequent. Table~\ref{tab:beaufort_detailed} in the appendix provides forecast error metrics across Beaufort wind-speed categories for reference.

\subsection{Spatiotemporal Analysis}
\paragraph{Lead time analysis.}
As illustrated in Fig.~\ref{fig:forecast-rmse-lead-time-obs}, RMSE increases gradually with lead time for all three models, from approximately 2.7–2.8 m\,s$^{-1}$ at +6 h to 3.1–3.2 m\,s$^{-1}$ at +72 h. HRES consistently achieves the lowest RMSE at every lead time, demonstrating a clear advantage over FCN3 and GraphCast throughout the forecast horizon. FCN3 and GraphCast perform similarly at shorter lead times, although FCN3 shows a notable increase after +48 h and records the highest RMSE at +72 h. The overlapping standard error bands show that the differences in mean RMSE between the models are small compared with the uncertainty across stations.

\paragraph{Station-based analysis.}
Shown in Fig.~\ref{fig:panel_rmse_station}, station-wise comparison reveals relatively similar RMSE values at most stations, but with larger differences found in a few locations. Bardufoss has the lowest RMSE for all three models, with values below 1.8 m\,s$^{-1}$. In contrast, Fakken is the most challenging station, where RMSE reaches 4.58 m\,s$^{-1}$ for FCN3, 5.15 m\,s$^{-1}$ for Graphcast, and 4.69 m\,s$^{-1}$ for HRES. Relatively high errors are also observed at Hekkingen Fyr, while most other stations have RMSE values between approximately 2.0 m\,s$^{-1}$ and 3.3 m\,s$^{-1}$. We provide additional station-level metrics as heatmaps in Appendix~\ref{sec:station-heatmap}. The relative performance of the three models varies across stations. HRES achieves the lowest RMSE at several sites, while FCN3 performs best at Tromsø--Langnes and Fakken, and GraphCast at Bardufoss, Torsvåg Fyr, and Banak. GraphCast also shows the largest error at Fakken. Despite these differences in ranking, the three models exhibit broadly similar station-wise error patterns, with relatively small differences at most stations.
\paragraph{Terrain-based analysis.}
We classify stations by geographical setting, including fjords, marine coasts, coastal ridges, and valleys, reflecting differences in surrounding terrain and wind exposure. Full station characteristics are provided in Appendix~\ref{sec:weather_station_info}. As illustrated in Fig.~\ref{fig:panel_rmse_station}, marine coast stations generally had lower RMSE than fjord and coastal ridge stations. Fakken, categorized as a coastal ridge, had the largest errors across all three models, whereas Bardufoss, located in a valley, had the lowest. Hekkingen Fyr also showed relatively high errors within the coastal ridge group. Andøya performed best among the marine coast stations, with RMSE ranging from 2.09 to 2.34 m\,s$^{-1}$. The fjord stations showed moderate errors, with Måsvik performing best, followed by Sørkjosen Lufthavn, while Banak consistently had the largest errors, ranging from 3.02 to 3.35 m\,s$^{-1}$.

\begin{table*}[tb]
\centering
\caption{High-wind performance across all stations during 2016--2022.}
\label{tab:high-wind-performance}
\begin{threeparttable}
\begin{tabular}{lrrrrcc}
\multicolumn{7}{r}{$N=40{,}247$} \\
\toprule
Model
& RMSE $\downarrow$
& nRMSE $\downarrow$
& Bias $(\to 0)$
& nBias $(\to 0)$ & Corr. & ACC \\
\midrule

HRES
& $6.05\,(1.88)$
& $45.59\,(14.97)$
& $-5.50\,(2.07)$
& $-41.51\,(16.69)$ & $.480\,(.156)$ & $.529\,(.401)$\\

\textbf{FCN3}
& $\mathbf{5.77\,(1.96)}$
& $\mathbf{43.42\,(15.45)}$
& $\mathbf{-4.97\,(2.39)}$
& $\mathbf{-37.58\,(18.75)}$ & $.464\,(.121)$ & $\mathbf{.554\,(.398)}$\\

GC
& $6.01\,(1.93)$
& $45.14\,(14.75)$
& $-5.43\,(2.16)$
& $-40.82\,(16.58)$  & $\bm{.488\,(.120)}$ & $.539\,(.413)$\\

\bottomrule
\end{tabular}
*\textbf{Bold} indicates the best result for each reference, and values are reported as the mean with the standard deviation given in parentheses. RMSE and Bias are in \(\mathrm{m\,s^{-1}}\), while nRMSE and nBias are in \%. GC: GraphCast.
\end{threeparttable}
\end{table*}

\subsection{Comparison of Training and Non-Training Years}
Recent work has reported temporal drift in boreal winter land temperature predictions, resembling climates from 15--20 years earlier than the prediction period~\citep{landsberg2026forecasting}. FCN3 was trained on ERA5 data from 1980--2016, tested on 2017 data, and validated out of sample over 2018--2021~\citep{Bonev2025FourCastNet3A}. For GraphCast, it was trained on ERA5 data from 1979--2017~\citep{lam2023learning}. Therefore, we compare the annual RMSE of the models to examine temporal performance changes within and outside their respective training periods. All three models showed their highest errors in 2020, followed by lower errors in 2021 and 2022, as presented in Fig.~\ref{fig:panel_rmse_year}. RMSE varied between years but showed no consistent increase over the post-training period. Similar variations across MLWP models and HRES suggest differences in annual forecast difficulty rather than systematic drift.
\begin{table}[ht]
\centering
\caption{Yearly statistical significance of RMSE differences relative to HRES.}
\label{tab:yearly_significance}
\small
\setlength{\tabcolsep}{4pt}
\begin{tabular}{@{}lllcc@{}}
\toprule
\textbf{Model} & \textbf{Year} &
\textbf{$\Delta$RMSE} & \textbf{95\% CI} &
\textbf{$p$} \\
\midrule
\multirow{7}{*}{FCN3}
& 2016 & +0.023 & [-0.016, 0.060] & 0.243 \\
& 2017 & +0.033 & [-0.021, 0.083] & 0.215 \\
& 2018 & +0.112 & [0.082, 0.142] & $<0.0001^{*}$ \\
& 2019 & +0.110 & [0.073, 0.147] & $<0.0001^{*}$ \\
& 2020 & +0.076 & [0.040, 0.112] & $<0.0001^{*}$ \\
& 2021 & +0.066 & [0.026, 0.105] & $0.001^{*}$ \\
& 2022 & +0.092 & [0.049, 0.136] & $<0.001^{*}$ \\
\midrule
\multirow{7}{*}{GC}
& 2016 & +0.018 & [-0.014, 0.047] & 0.244 \\
& 2017 & +0.028 & [-0.016, 0.069] & 0.203 \\
& 2018 & +0.074 & [0.050, 0.098] & $<0.0001^{*}$ \\
& 2019 & +0.056 & [0.019, 0.094] & $0.004^{*}$ \\
& 2020 & +0.044 & [0.007, 0.079] & $0.018^{*}$ \\
& 2021 & +0.027 & [-0.016, 0.067] & 0.201 \\
& 2022 & +0.065 & [0.039, 0.092] & $<0.0001^{*}$ \\
\bottomrule
\addlinespace[3pt]
\multicolumn{5}{@{}l}{\footnotesize GC: GraphCast, and $^{*}$indicates statistical significance.} \\
\end{tabular}
\end{table}
To assess temporal generalization, we analyzed yearly RMSE differences relative to HRES and extended MLWP inference to 2010 to compare training and post-training periods. We estimated 95\% confidence intervals using a temporal block bootstrap with 7-day blocks~\citep{lahiri2013resampling, buhlmann2002bootstraps, hongyi1996bootstrapping}. As shown in Table~\ref{tab:yearly_significance}, the relative RMSE differences fluctuate without a consistent increasing trend. Furthermore, neither FCN3 nor GraphCast shows statistically significant RMSE differences between training and post-training periods across the tested block lengths (Table~\ref{tab:block_sensitivity}), providing no statistically significant evidence of systematic post-training performance degradation over the evaluated periods.

\subsection{High-Wind Analysis}
Under high-wind conditions, all models substantially underestimate the observed wind speeds, as indicated by larger negative biases and RMSE values around twice as high as those in the overall evaluation from Table~\ref{tab:high-wind-performance}, with normalized RMSE ranging from -43.42\% to -45.59\%. FCN3 performs best in this regime, achieving the lowest RMSE and nRMSE as well as the smallest absolute bias. HRES shows the largest errors and strongest underestimation, while GraphCast consistently ranks between the two models. However, it is worth noting that the comparison may be affected by HRES being evaluated at 0.25° rather than its native 0.1° resolution to ensure a fair comparison of spatial resolution, as well as differences in initial conditions between HRES and the MLWP models. The statistical significance results under high-wind conditions are provided in Appendix~\ref{app:highwind_significance}.

\begin{figure}[ht]
    \centering
    \begin{subfigure}[t]{0.48\linewidth}
        \centering
        \includegraphics[width=\linewidth]
        {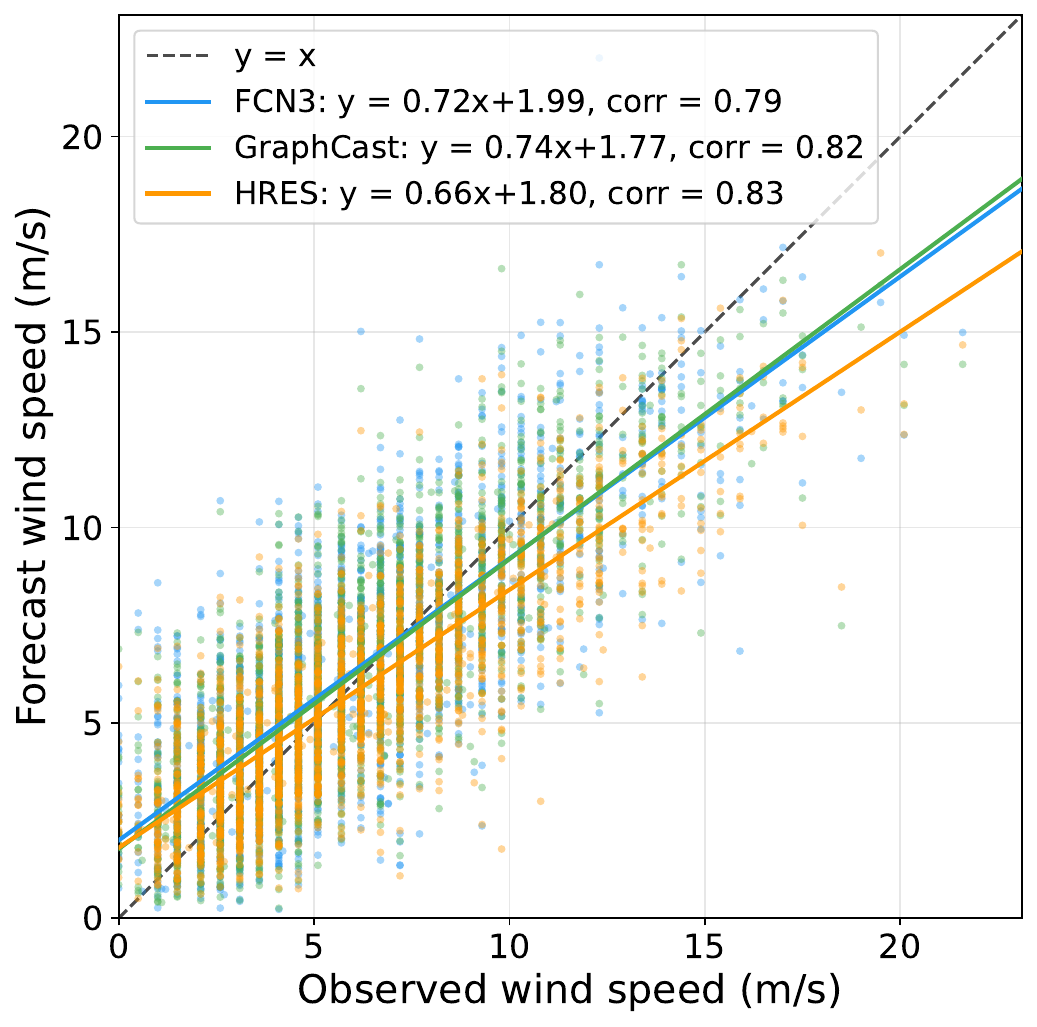}
        \caption{Andøya}
        \label{fig:high_wind_andoya}
    \end{subfigure}
    \hfill
    \begin{subfigure}[t]{0.48\linewidth}
        \centering
        \includegraphics[width=\linewidth]
        {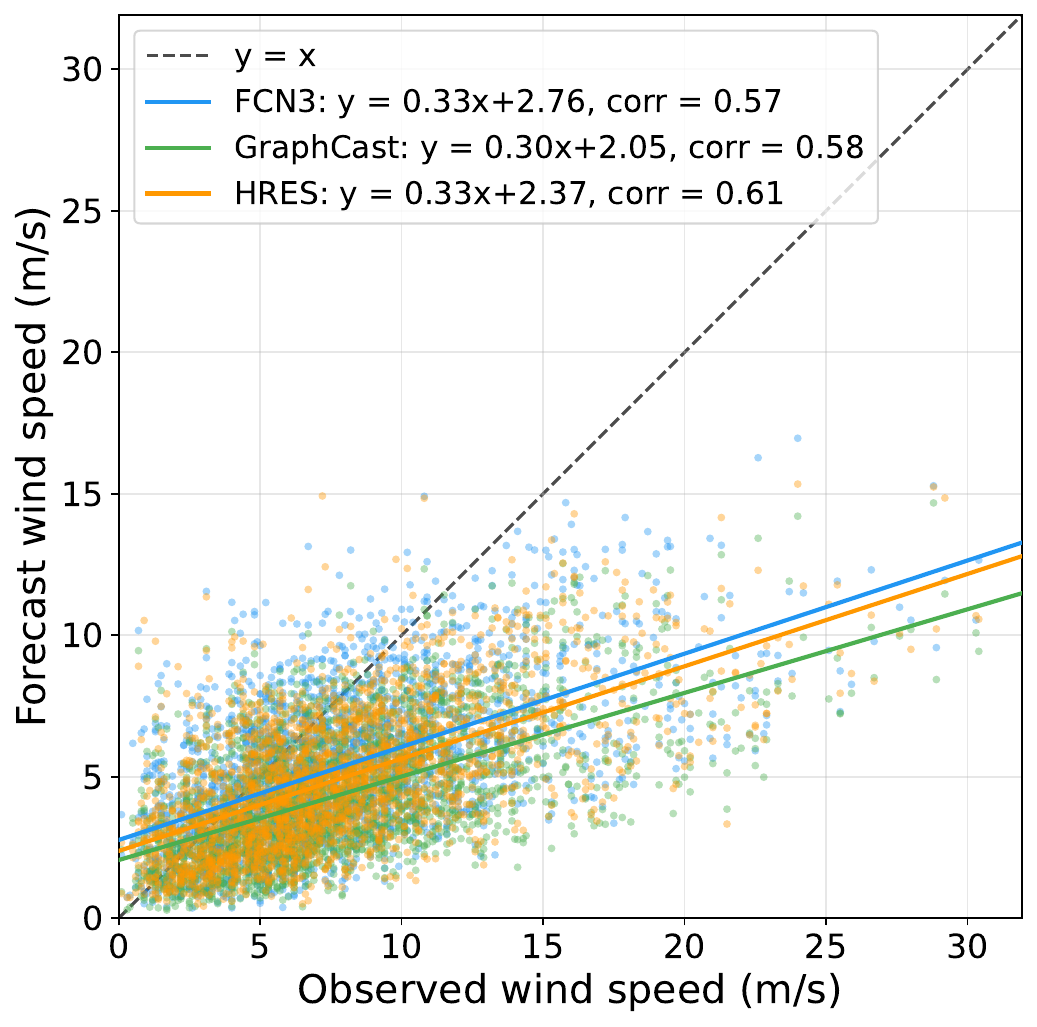}
        \caption{Fakken}
        \label{fig:high_wind_fakken}
    \end{subfigure}

    \vspace{0.3em}

    \begin{subfigure}[t]{0.48\linewidth}
        \centering
        \includegraphics[width=\linewidth]
        {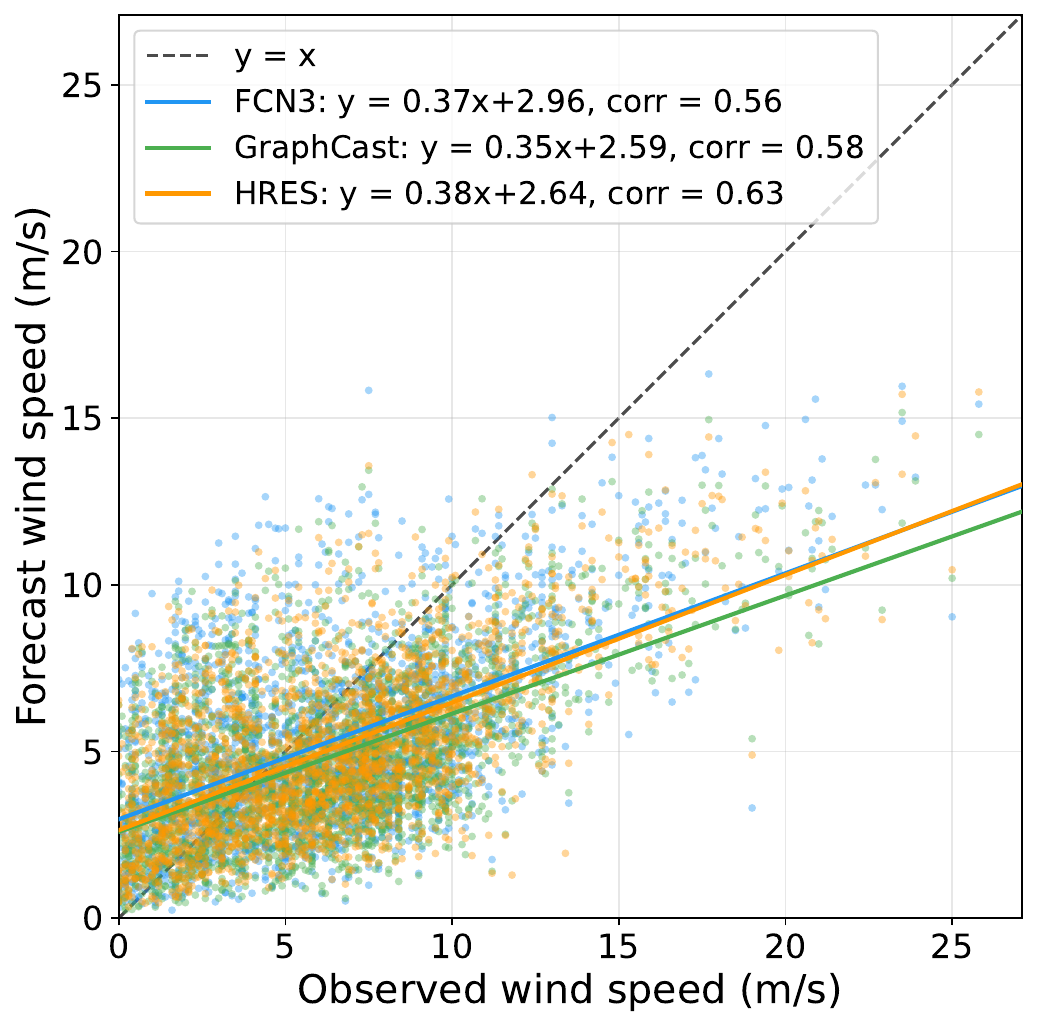}
        \caption{Hekkingen Fyr}
        \label{fig:high_wind_hekkingen}
    \end{subfigure}
    \hfill
    \begin{subfigure}[t]{0.48\linewidth}
        \centering
        \includegraphics[width=\linewidth]
        {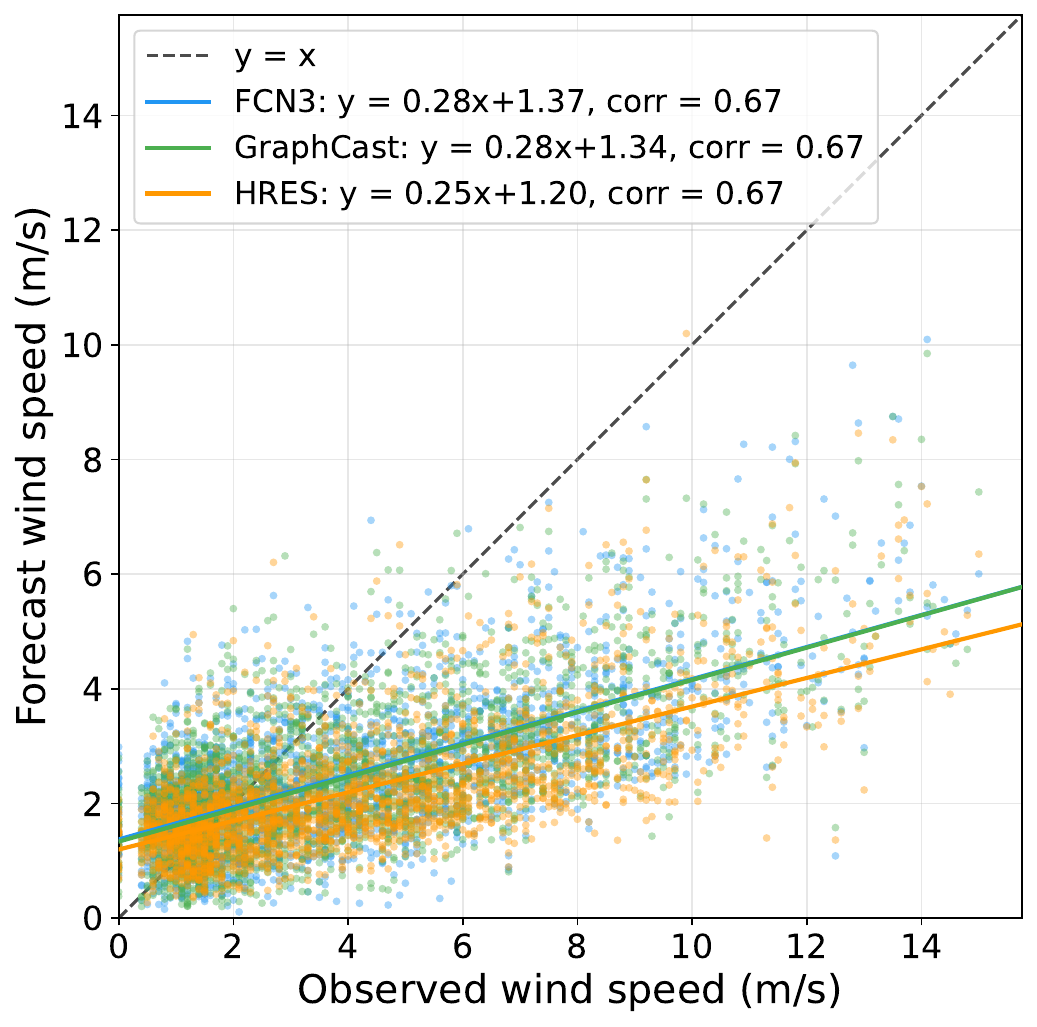}
        \caption{Tromsø--Langnes}
        \label{fig:high_wind_tromso}
    \end{subfigure}

    \caption{Scatter plots of observed versus predicted wind speeds at four selected stations.}
    \label{fig:high_wind_selected_stations}
\end{figure}

We show scatter plots of observed versus predicted wind speeds at +24 h in Fig.~\ref{fig:high_wind_selected_stations}, and find notable differences across four representative stations. Andøya had the best performance, with correlations of 0.79--0.83 and slopes of 0.66--0.72, indicating that all models captured both the variation and magnitude of wind speed relatively well. However, performance was much weaker at Fakken, Hekkingen Fyr, and Tromsø--Langnes, where the slopes ranged from only 0.25 to 0.38. These low slopes indicate that the predicted wind speed range was strongly compressed, leading to increasing underestimation as the observed wind speed became higher. Although HRES generally achieved the highest correlation, its slopes were not consistently closest to one. Overall, station-to-station differences were larger than differences among models, and all three models struggled to reproduce the strong observed winds. Scatter plots for the remaining stations are provided in the Appendix~\ref{sec:scatter-plots}.
\subsection{Computational Cost}
We compare computational costs in terms of training, inference time, and memory usage across forecast lengths. As pretrained FCN3 and GraphCast checkpoints were used without retraining, we report their original training costs. FCN3 required 78 h on 1,024 H100 GPUs, 15 h on 512 A100 GPUs, and 8 h on 256 H100 GPUs across three stages, while GraphCast required approximately four weeks on 32 TPU v4 devices~\citep{lam2023learning,bodnar2025foundation}. HRES does not involve an equivalent offline training stage.
For inference, both models were evaluated on a single NVIDIA H200 GPU with 4 CPU cores. At 0.25° resolution, FCN3 required 90.37 ± 1.28 s per 72-h forecast, compared with 139.90 ± 1.98 s for GraphCast. HRES inference was not measured directly, as precomputed WeatherBench 2 forecasts were used. For reference, ECMWF reports approximately 30 minutes for a 10-day HRES forecast at 0.1° resolution on 128 HPC nodes~\citep{geer2025data}. Across 3–15-day forecasts (Table~\ref{tab:inference_cost}), inference time increased from 95.93 to 219.94 s for FCN3 and from 151.43 to 247.05 s for GraphCast. FCN3 was faster but consumed more GPU and CPU memory. For details on additional computational resources used in this study, see Appendix~\ref{sec:computional-resources}

\section{Discussion}
Recent advances in MLWP models have demonstrated competitive or superior forecast skill compared with traditional NWP systems~\citep{bi2023accurate,lam2023learning,chen2023fuxi}. Our results suggest that the advantages reported in global and grid-based benchmark evaluations may not be consistently observed when local near-surface wind forecasts are evaluated against station observations in complex terrain. Although FCN3 and GraphCast outperformed HRES when compared against ERA5, this is not unexpected given that both models were trained on ERA5 data~\citep{Bonev2025FourCastNet3A,lam2023learning}. A complementary evaluation using ERA5 as the reference is provided in Appendix~\ref{sec:eval-era5}. 

Beyond the choice of reference, the station-level evaluation also revealed pronounced spatial variability in forecast performance, which may partly reflect differences in local geography and exposure. Bardufoss, for example, is a valley sheltered by surrounding mountains, where weak winds and frequent wintertime inversions contribute to low RMSE. However, global MLWP models with coarse resolution may predict low winds because the grid cell is represented as mountainous terrain by assuming strong surface friction, resulting in a low RMSE without necessarily capturing the actual local wind dynamics. In addition, the models are penalized at fjord stations such as Tromsø-Langnes, Banak, and Sørkjosen Lufthavn, where the compound effects of terrain and surface characteristics are massively different from what the model assumes. Furthermore, in the area of the coastal ridge like Fakken or Hekkingen Fyr, strong wind channelling and downslope wind storm effects can produce larger forecast errors that remain unresolved by coarse model grids. 

Our central finding is that HRES slightly outperforms FCN3 and GraphCast at station-level evaluation, but FCN3 and GraphCast remain comparable to HRES over Northern Norway. The differences are less than 0.07~m~s$^{-1}$ and are considerably smaller than the variation observed among stations. Therefore, the aggregate results support a modest numerical advantage for HRES rather than a general conclusion that HRES is consistently superior. It is worth noting that our main analysis focuses on the three global forecasting models. For additional context, we also evaluated ERA5 and the regional CARRA1~\citep{copernicus2021arctic} reanalysis against station observations to examine differences between reanalysis products and local observations, with the corresponding results provided in Appendix~\ref{sec:comp-era5-carra1-obs}.

\paragraph{Limitations and future work.} Several limitations should be considered in this case study. First, the evaluation is based on only 12 stations in Northern Norway, and the findings may therefore not extend directly to other regions with different climates, terrain, or observation networks. Future work could evaluate more regions with similarly complex coastal and mountainous terrain to assess the generalizability of the conclusions beyond Northern Norway. Second, bilinear interpolation provides only an approximate mapping from gridded forecasts to station locations. Although widely used recently, it cannot fully represent local effects associated with elevation, coastline geometry, surface roughness, and unresolved terrain, and the resulting errors may therefore reflect both forecast error and station-to-grid representativeness error~\citep{ben2020accounting,gober2008could,janjic2018representation}. Future improvements may therefore require higher-resolution regional models or downscaling methods that explicitly incorporate local topography and surface characteristics. 

\section{Conclusion}
We compared the performance of two MLWP models, FCN3 and GraphCast, with the ECMWF HRES model for station-level wind forecasts in Northern Norway. Overall, HRES achieved the lowest errors (2.89 m\,s$^{-1}$), while FourCastNet3 (2.96 m\,s$^{-1}$) and GraphCast (2.94 m\,s$^{-1}$) delivered comparable results over the evaluation period. We found that errors increased with lead time, but spatial differences between stations were more pronounced. Under high-wind conditions, all models became less accurate and significantly underestimated wind speeds, as indicated by the increased RMSE and more negative bias. Our findings of this study highlight important directions for guiding future work, such as improving the representation of complex geography and the prediction of high-wind events in machine learning-based weather forecasting.

\section*{Acknowledgments}
This work was funded and supported by The Research Council of Norway (Visual Intelligence grant no. 309439 and KnowEarth project no. 337481). We acknowledge Sigma2 – the National Infrastructure for High Performance Computing and Data Storage in Norway, for providing computational resources to this project through grant no. NN8106K.

\section*{LLM Disclosure}
In this work, we used generative AI tools to assist with software development, literature research, and manuscript editing. Specifically, we used ChatGPT 5.6 Sol via the web interface to identify relevant prior work and literature on MLWP models. All references were then checked by the authors against original sources. Additionally, we used the ChatGPT 5.6 series to improve clarity and readability of the manuscript, primarily by helping to shorten the existing text. The authors reviewed and revised all AI-assisted text. Finally, Claude Code (Claude Opus 5.5) was also used to assist with code implementation, especially environment setup and data loading. All AI-assisted code was subsequently reviewed and validated through testing by the authors before its use in the experiments. We take responsibility for the final content of this work, including text, claims, or artifacts produced with the aid of LLMs/AI tools.

\printbibliography
\appendix
\section{Computational Resources}\label{sec:computional-resources}
\begin{table}[ht]
\centering
\small
\caption{Computational requirements for the FCN3 and GraphCast experiments. Execution times and peak memory usage are reported for individual annual tasks.}
\label{tab:computational_resources}
\begin{tabular}{lcc}
\toprule
Configuration & Time (hours) & CPU Memory (GB) \\
\midrule
FCN3 & 9.05--9.50 & 66--68 \\
GraphCast & 13.99--14.51 & 58--61 \\
\bottomrule
\end{tabular}
\end{table}
All FCN3 and GraphCast inference experiments in this study were conducted on the Olivia GPU cluster provided by the Norwegian Research Infrastructure Services (NRIS)~\citep{sigma2_olivia}. Each experimental configuration was submitted as a SLURM job array containing seven tasks, with one task assigned to each year from 2016 to 2022. Each task requested one GPU, four CPU cores, 192~GB of memory, and a maximum execution time of 40 hours. Subject to GPU availability, the seven annual tasks could run concurrently, making the completion time of each configuration approximately equal to that of its slowest annual task. FCN3 and GraphCast were each executed as a SLURM array containing seven parallel tasks, with one task corresponding to each evaluation year. Each FCN3 task required approximately 9.05--9.50 hours, while each GraphCast task required approximately 13.99--14.51 hours. Because the annual tasks were executed in parallel, the wall clock time for each model was determined by its slowest task, as shown in Table~\ref{tab:computational_resources}. Across the fourteen annual tasks, the two experiments consumed approximately 163 GPU hours in total, and the forecast outputs and analysis data were stored on the cluster work filesystem. For comparison, a 10-day HRES forecast requires about 30 minutes on 128 HPC nodes, whereas GraphCast produces a 10-day forecast in about 1 minute on a single TPU~\citep{geer2025data}. Therefore, assuming approximately linear scaling with forecast length, our 72-h forecast horizon would correspond to about 9 minutes per HRES forecast and roughly 384 HPC-node wall-clock hours over the same evaluation period. Additionally, Table~\ref{tab:inference_cost} presents the inference times of FCN3 and GraphCast across different forecast lengths, along with their GPU and CPU memory usage.
\begin{table}[htbp]
\centering
\caption{Inference time and memory usage of FCN3 and GraphCast.}
\label{tab:inference_cost}
\small
\setlength{\tabcolsep}{3pt}
\begin{tabular}{@{}lllllll@{}}
\toprule
\textbf{Model} & \textbf{3d} & \textbf{5d} &
\textbf{10d} & \textbf{15d} &
\textbf{GPU} & \textbf{CPU} \\
\midrule
FCN3 & 95.93 & 109.39 & 164.59 & 219.94 & 52.66 & 65.2 \\
GC & 151.43 & 157.03 & 199.48 & 247.05 & 38.00 & 59.1 \\
\bottomrule
\end{tabular}
\parbox{\columnwidth}{\scriptsize
*Inference time is reported in seconds (s), and memory usage in GB. Times are averaged over three initializations for each forecast length. GC: GraphCast
}
\end{table}
\newpage
\section{Beaufort Force}\label{sec:beaufort-force}
The Beaufort force scale provides a categorical description of wind intensity based on wind speed. Table~\ref{tab:beaufort_scale} summarizes the Beaufort force categories and their corresponding wind-speed ranges used in this study.
\begin{table}[tbhp]
\centering
\caption{Beaufort wind force scale.}
\label{tab:beaufort_scale}
\begin{tabular}{clc}
\toprule
\textbf{Force} & \textbf{Description} & \textbf{Wind speed (m/s)} \\
\midrule
0  & Calm            & 0.0--0.5    \\
1  & Light air       & 0.5--1.6    \\
2  & Light breeze    & 1.6--3.4    \\
3  & Gentle breeze   & 3.4--5.5    \\
4  & Moderate breeze & 5.5--8.0    \\
5  & Fresh breeze    & 8.0--10.8   \\
6  & Strong breeze   & 10.8--13.9  \\
7  & Near gale       & 13.9--17.2  \\
8  & Gale            & 17.2--20.8  \\
9  & Strong gale     & 20.8--24.5  \\
10 & Storm           & 24.5--28.5  \\
11 & Violent storm   & 28.5--32.7  \\
12 & Hurricane       & $\geq$32.7  \\
\bottomrule
\end{tabular}
\end{table}
\section{Temporal and Shared Weather-Event Dependence Analysis}\label{app:block_bootstrap}
A complementary moving block bootstrap~\citep{lahiri2013resampling, buhlmann2002bootstraps, hongyi1996bootstrapping} was performed to account for temporal correlation and shared weather events across stations. Specifically, consecutive calendar days were resampled in blocks of 1, 3, 7, and 14 days, keeping all stations and forecast lead times together within each block. Each bootstrap sample contains approximately 366,585 high-wind samples per model. Table~\ref{tab:block_bootstrap} presents the pairwise RMSE differences and their 95\% confidence intervals for different temporal block lengths. All pairwise RMSE differences remained statistically significant across block lengths (95\% confidence intervals). Despite this statistical significance, the absolute RMSE differences remain small, indicating broadly comparable overall performance among the three models. However, we also acknowledge that the effective number of spatial clusters remained limited to 12 stations.
\begin{table}[ht]
\centering
\caption{Moving block bootstrap results for pairwise RMSE differences
using different temporal block lengths.}
\label{tab:block_bootstrap}
\begin{tabular}{llll}
\toprule
Comparison & $\Delta$RMSE & Block & 95\% CI \\
\midrule
\multirow{4}{*}{FCN3-GC}
 & +0.029 & 1 day   & [.019, .038] \\
 & +0.029 & 3 days  & [.018, .040] \\
 & +0.029 & 7 days  & [.016, .041] \\
 & +0.029 & 14 days & [.015, .043] \\
\midrule
\multirow{4}{*}{FCN3-HRES}
 & +0.073 & 1 day   & [.062, .085] \\
 & +0.073 & 3 days  & [.059, .087] \\
 & +0.073 & 7 days  & [.057, .088] \\
 & +0.073 & 14 days & [.055, .091] \\
\midrule
\multirow{4}{*}{GC-HRES}
 & +0.044 & 1 day   & [.034, .055] \\
 & +0.044 & 3 days  & [.032, .057] \\
 & +0.044 & 7 days  & [.031, .058] \\
 & +0.044 & 14 days & [.029, .058] \\
\bottomrule
\end{tabular}

\smallskip
\footnotesize{GC: GraphCast. All 95\% confidence intervals exclude zero, meaning statistically significant pairwise RMSE differences.}
\end{table}
\section{Weather Stations}
\label{sec:weather_station_info}
\begin{figure*}[t]
    \centering

    \begin{subfigure}[t]{0.49\textwidth}
        \centering
        \includegraphics[width=\linewidth]{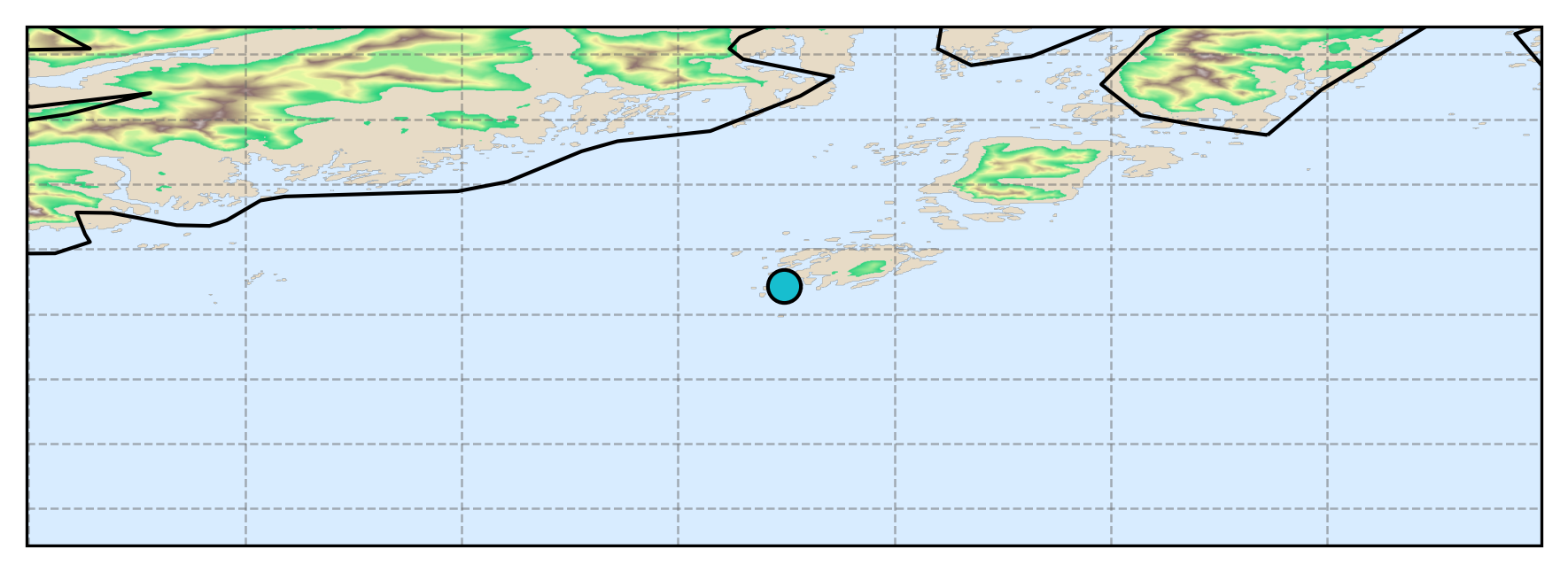}
        \caption{Skrova Fyr (SN85380)}
    \end{subfigure}
    \hfill
    \begin{subfigure}[t]{0.49\textwidth}
        \centering
        \includegraphics[width=\linewidth]{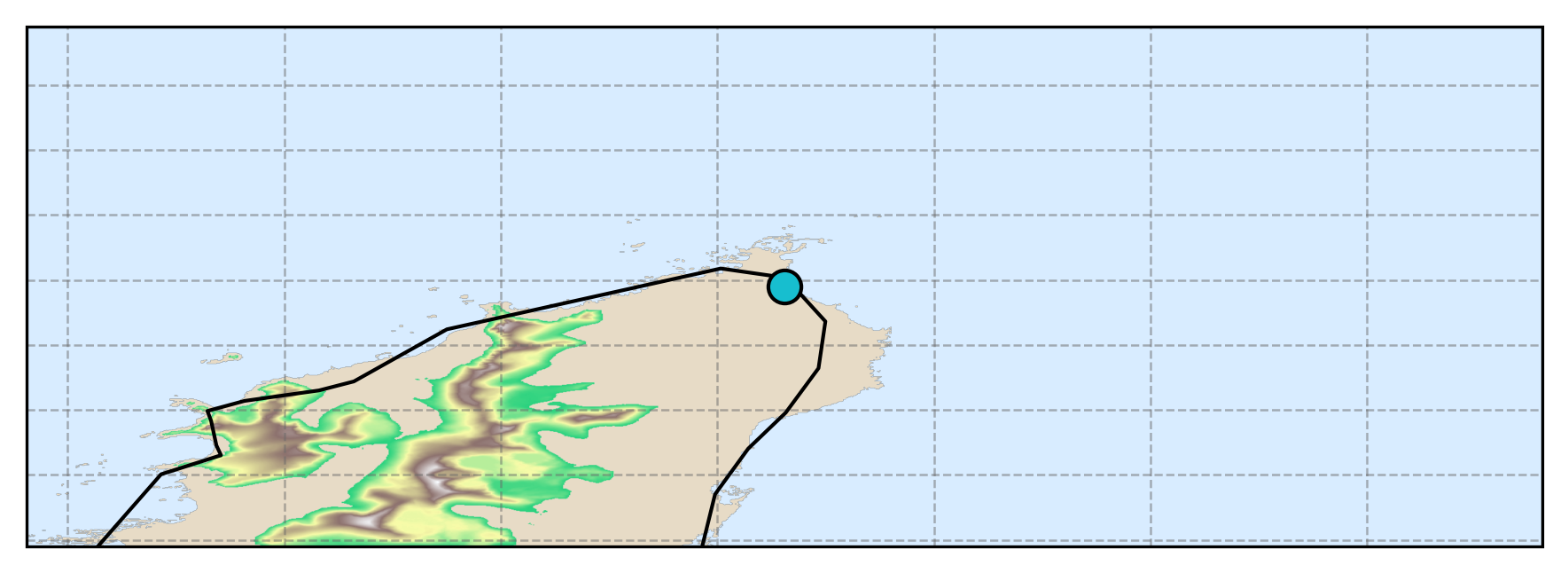}
        \caption{Andøya (SN87110)}
    \end{subfigure}

    \begin{subfigure}[t]{0.49\textwidth}
        \centering
        \includegraphics[width=\linewidth]{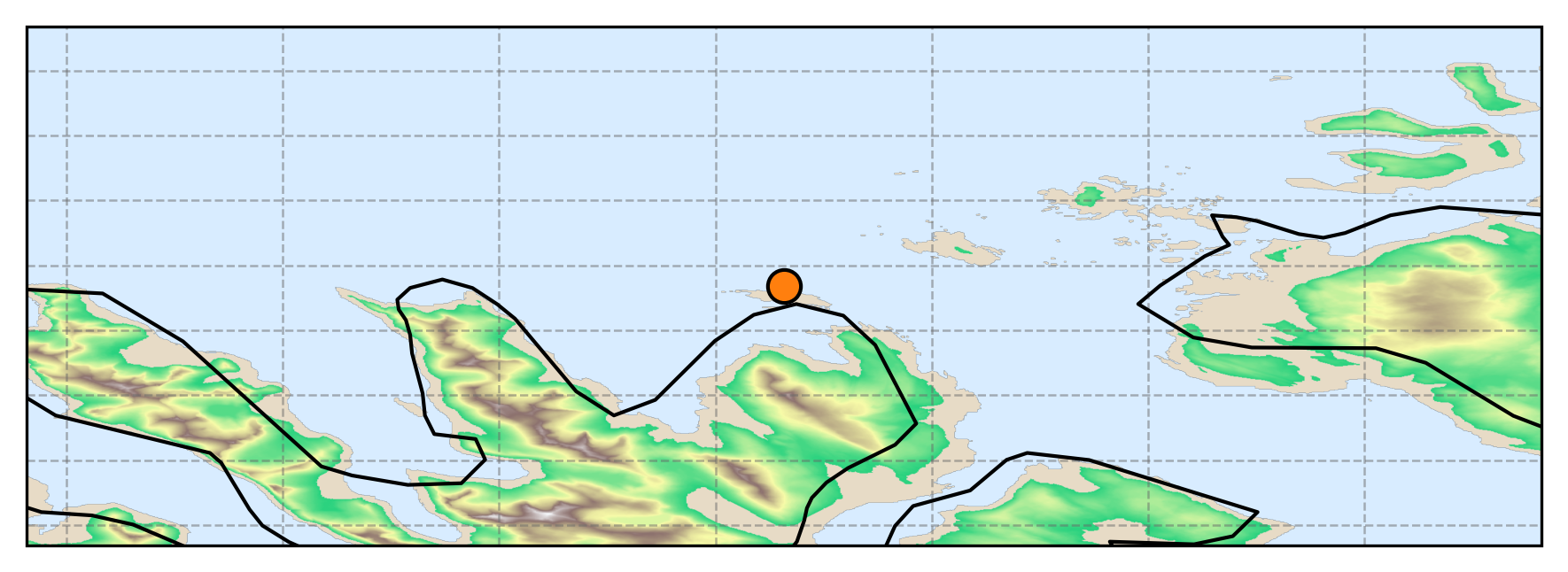}
        \caption{Hekkingen Fyr (SN88690)}
    \end{subfigure}
    \hfill
    \begin{subfigure}[t]{0.49\textwidth}
        \centering
        \includegraphics[width=\linewidth]{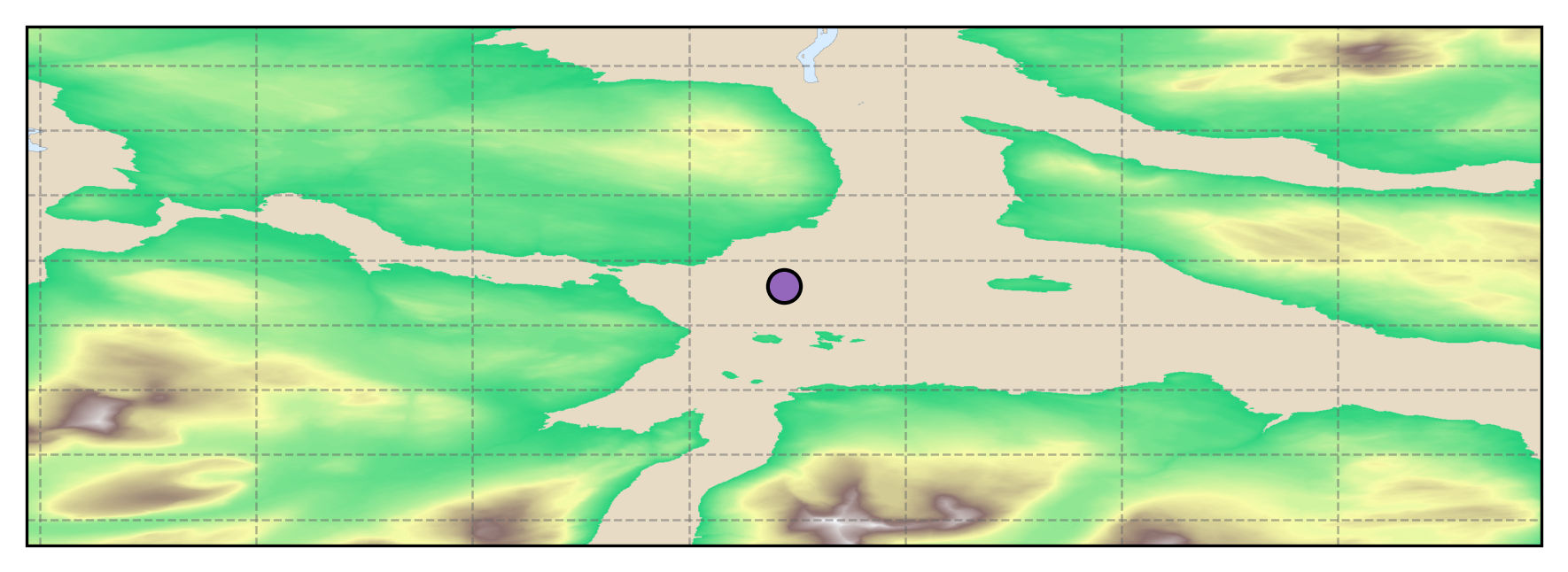}
        \caption{Bardufoss (SN89350)}
    \end{subfigure}

    \begin{subfigure}[t]{0.49\textwidth}
        \centering
        \includegraphics[width=\linewidth]{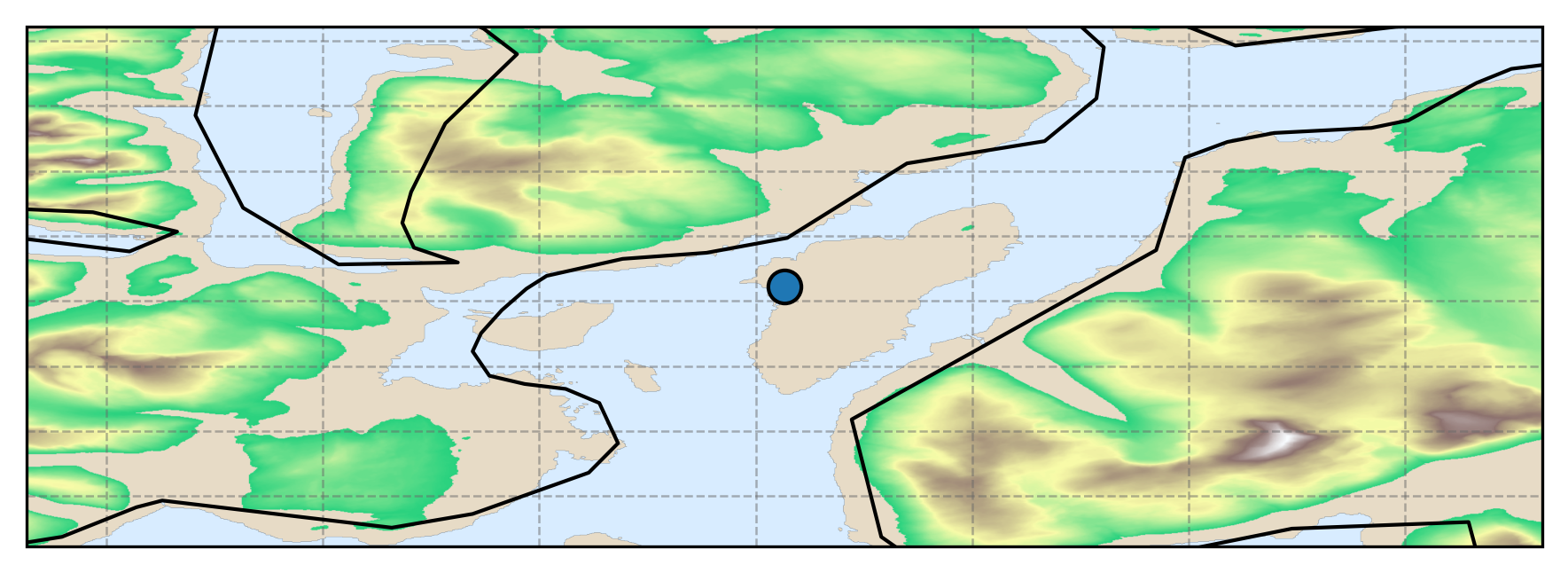}
        \caption{Tromsø-Langnes (SN90490)}
    \end{subfigure}
    \hfill
    \begin{subfigure}[t]{0.49\textwidth}
        \centering
        \includegraphics[width=\linewidth]{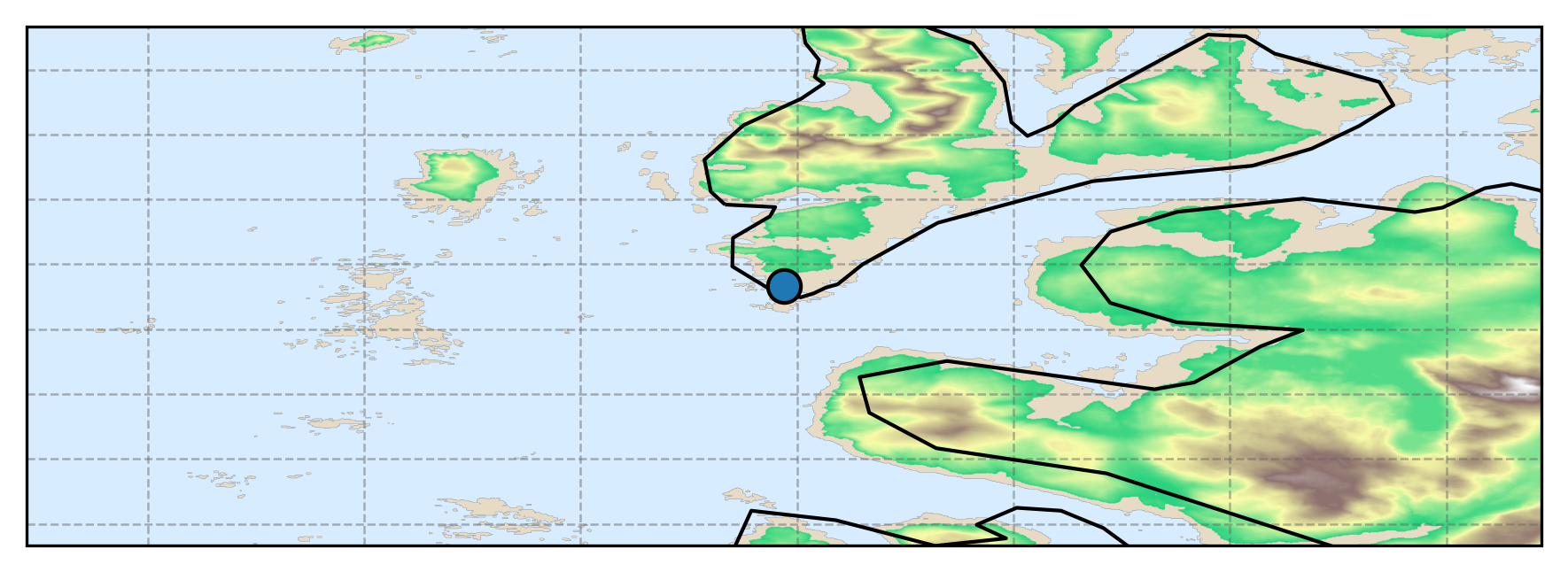}
        \caption{Måsvik (SN90720)}
    \end{subfigure}

    \begin{subfigure}[t]{0.49\textwidth}
        \centering
        \includegraphics[width=\linewidth]{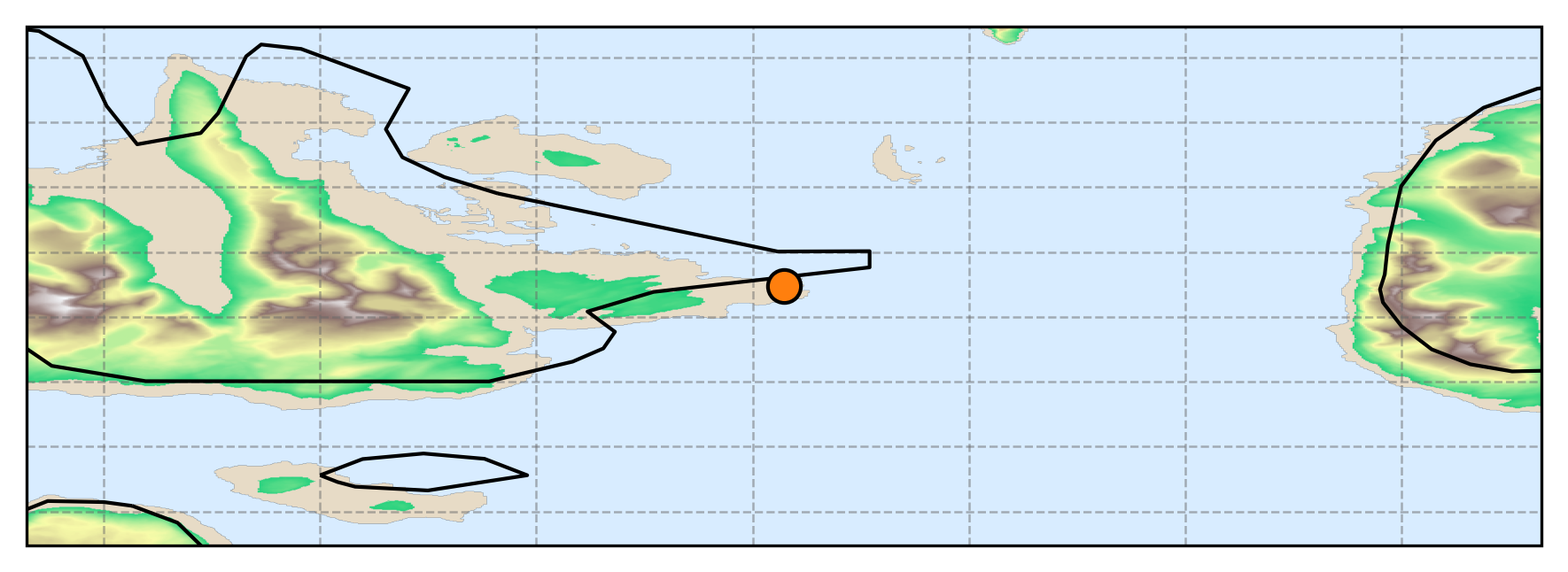}
        \caption{Fakken (SN90760)}
    \end{subfigure}
    \hfill
    \begin{subfigure}[t]{0.49\textwidth}
        \centering
        \includegraphics[width=\linewidth]{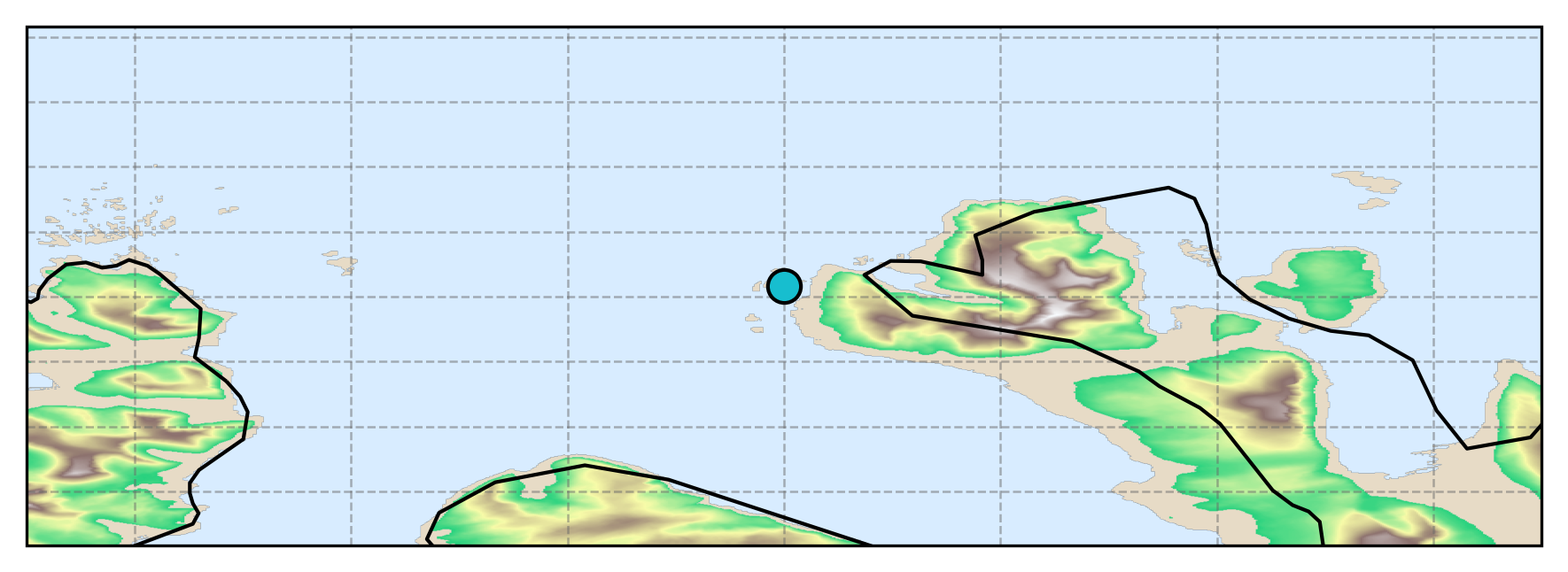}
        \caption{Torsvåg Fyr (SN90800)}
    \end{subfigure}

    \begin{subfigure}[t]{0.49\textwidth}
        \centering
        \includegraphics[width=\linewidth]{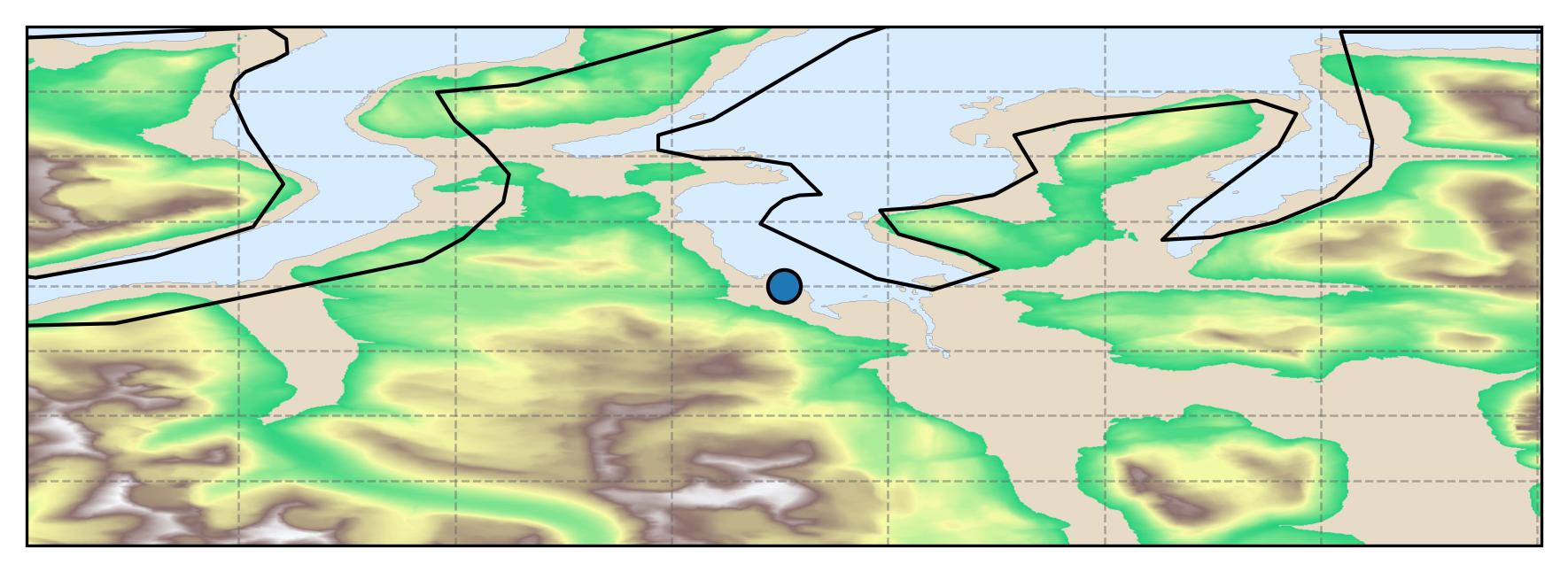}
        \caption{Sørkjosen Lufthavn (SN91740)}
    \end{subfigure}
    \hfill
    \begin{subfigure}[t]{0.49\textwidth}
        \centering
        \includegraphics[width=\linewidth]{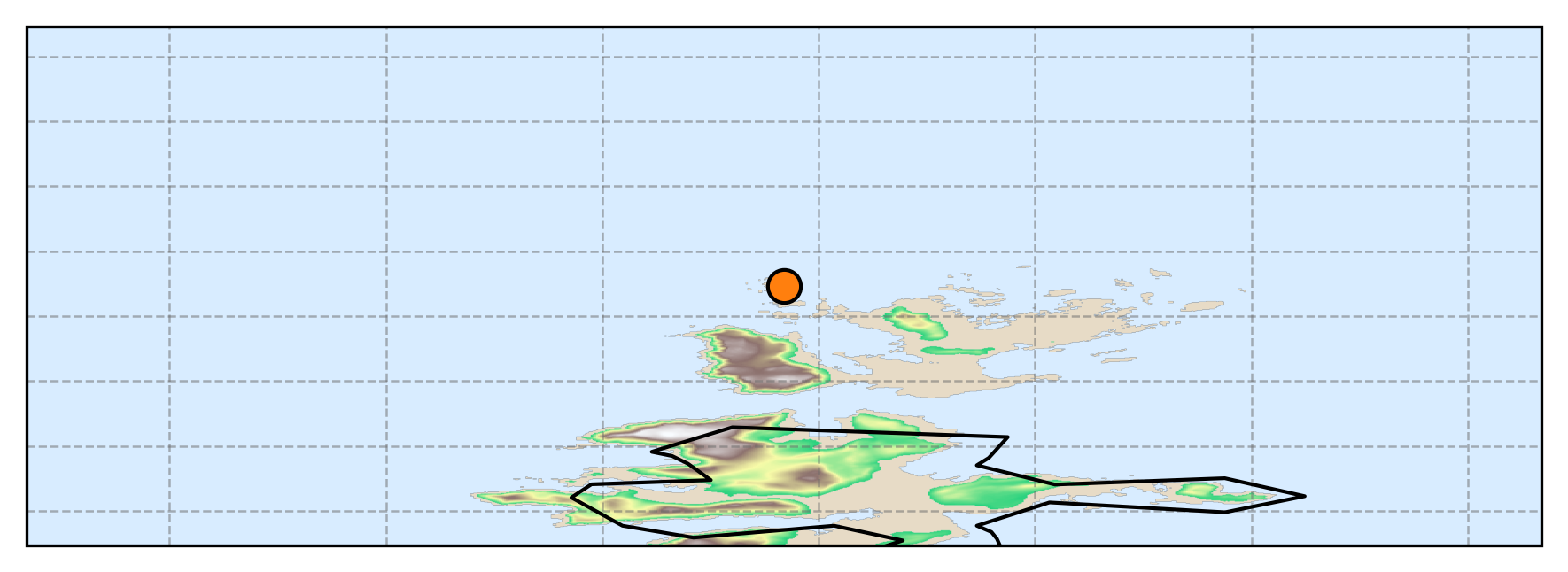}
        \caption{Fruholmen Fyr (SN94500)}
    \end{subfigure}

    \begin{subfigure}[t]{0.49\textwidth}
        \centering
        \includegraphics[width=\linewidth]{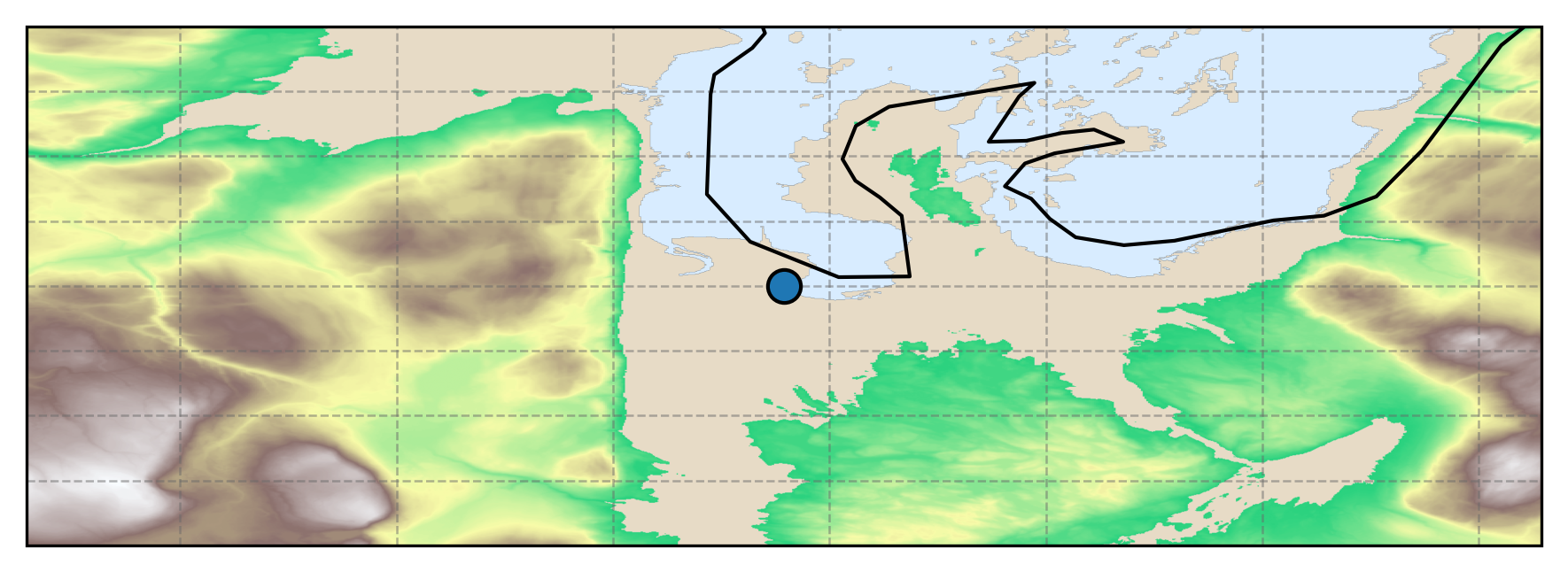}
        \caption{Banak (SN95350)}
    \end{subfigure}
    \hfill
    \begin{subfigure}[t]{0.49\textwidth}
        \centering
        \includegraphics[width=\linewidth]{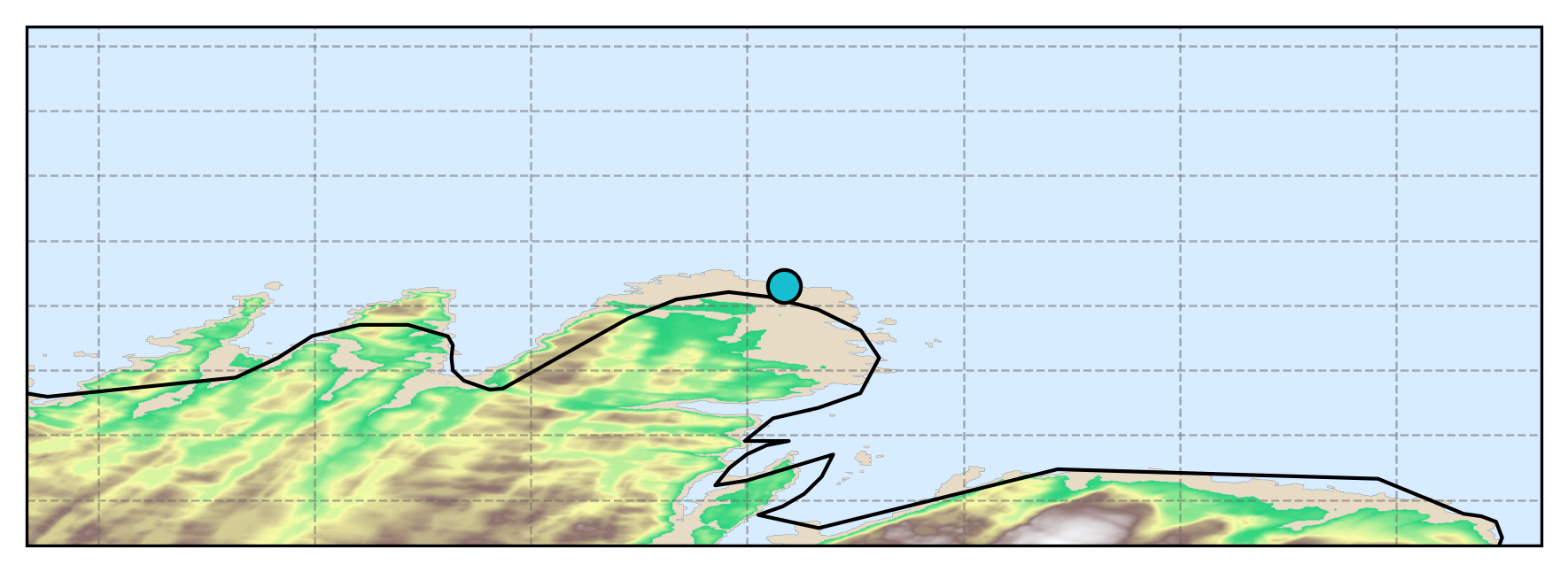}
        \caption{Slettnes Fyr (SN96400)}
    \end{subfigure}

    \caption{Zoomed-in views of the local surroundings of the 12 stations.}
    \label{fig:station_zoom_in}
\end{figure*}

Table~\ref{tab:station_characteristics} aggregates the geographical characteristics of the 12 observation stations used in the evaluation, including their coordinates, elevation, and terrain category. The stations cover a range of coastal, fjord, and inland valley environments across Northern Norway, with elevations ranging from near sea level to approximately 76~m. We also provide zoomed-in views of the local surroundings of the 12 stations in Fig.~\ref{fig:station_zoom_in}. These metadata provide context for interpreting station-level differences in forecast performance.
\begin{table*}[t]
\centering
\caption{Geographical and terrain characteristics of the 12 observation stations.}
\label{tab:station_characteristics}
\resizebox{\textwidth}{!}{
\begin{tabular}{llcccc}
\toprule
\textbf{Station ID} &
\textbf{Station} &
\textbf{Latitude} &
\textbf{Longitude} &
\textbf{Elevation} &
\textbf{Terrain label}\\
\midrule
SN91740 & Sørkjosen Lufthavn & 69.7900 & 20.9520 & 2.76  & Fjord \\
SN95350 & Banak              & 70.0600 & 24.9790 & 6.54  & Fjord \\
SN90720 & Måsvik             & 69.9900 & 18.6940 & 11.04 & Fjord \\
SN90490 & Tromsø-Langnes     & 69.6767 & 18.9133 & 8.97  & Fjord \\
SN87110 & Andøya             & 69.3070 & 16.1310 & 4.35  & Marine Coast \\
SN96400 & Slettnes Fyr       & 71.0890 & 28.2170 & 9.11  & Marine Coast \\
SN90800 & Torsvåg Fyr        & 70.2450 & 19.5000 & 18.30 & Marine Coast \\
SN85380 & Skrova Fyr         & 68.1530 & 14.6490 & 18.04 & Marine Coast \\
SN88690 & Hekkingen Fyr      & 69.6005 & 17.8317 & 28.51 & Coastal Ridge \\
SN94500 & Fruholmen Fyr      & 71.0940 & 23.9840 & 23.96 & Coastal Ridge \\

SN90760 & Fakken             & 70.1043 & 20.1145 & 56.49 & Coastal Ridge \\
SN89350 & Bardufoss          & 69.0580 & 18.5440 & 75.51 & Valley \\
\bottomrule
\end{tabular}
}
\end{table*}

\section{Evaluation against ERA5}\label{sec:eval-era5}
We additionally evaluated the three forecast models against ERA5 using $N=366{,}948$ samples. Table~\ref{tab:overall_metrics_era5} shows that all three models exhibited much lower RMSE values, ranging from 1.14  m\,s$^{-1}$ to 1.31 m\,s$^{-1}$. Among them, GraphCast achieved the lowest RMSE and nRMSE, while FCN3 demonstrated the smallest systematic bias, with an nBias of only 0.81\%. In contrast, GraphCast and HRES showed negative nBias values of -3.30\% and -4.95\%, respectively.

\begin{table*}[tbh]
\centering
\caption{Model performance evaluated against ERA5 across all stations during 2016--2022.}
\label{tab:overall_metrics_era5}
\setlength{\tabcolsep}{6pt}
\begin{threeparttable}
\begin{tabular}{clccccc}
\multicolumn{7}{r}{$N=366{,}948$} \\
\toprule
Model
& RMSE $\downarrow$
& nRMSE $\downarrow$
& Bias $(\to 0)$
& nBias $(\to 0)$
& Corr. & ACC \\
\midrule
HRES
& $1.31,(0.57)$
& $27.17,(4.93)$
& $-0.32,(0.58)$
& $-4.95,(9.46)$
& $.853(.036)$
& $.8238(.061)$\\
FCN3
& $1.24,(0.40)$
& $26.13,(2.67)$
& $\bm{0.04,(0.10)}$
& $\bm{0.81,(2.17)}$
& $.846(.036)$
& $.8365(.039)$\\
GC
& $\bm{1.14,(0.37)}$
& $\bm{24.34,(3.80)}$
& $-0.37,(0.46)$
& $-5.30,(9.49)$
& $\bm{.890(.034)}$
& $\bm{.8611(.036)}$\\
\bottomrule
\end{tabular}
*\textbf{Bold} indicates the best result, and values are reported as the mean with the standard deviation given in parentheses. RMSE and Bias are in $\mathrm{m\,s^{-1}}$, while nRMSE and nBias are in \%. GC: GraphCast
\end{threeparttable}
\end{table*}

\section{Comparison of CARRA1 and ERA5 with Station Observations}\label{sec:comp-era5-carra1-obs}
We further assess CARRA1 and ERA5 against station observations to examine how well the two reanalysis datasets represent local near-surface wind conditions. CARRA1 is a regional Arctic reanalysis based on the HARMONIE-AROME forecasting system and has a horizontal grid spacing of approximately 2.5~km~\citep{copernicus2021arctic}, substantially finer than the approximately 31~km resolution of ERA5~\citep{hersbach2020era5}. Its higher spatial resolution and improved representation of regional geography are designed to better resolve local atmospheric conditions in complex Arctic terrain and coastal regions. Previous evaluation over the north-east European Arctic has shown that CARRA generally agrees better with surface observations than ERA5, including for 10-m wind speed, although the magnitude of the added value varies spatially and seasonally~\citep{koltzow2022value}. Table~\ref{tab:carra1_era5} compares CARRA1 and ERA5 against the station observations. CARRA1 achieves a lower RMSE of 2.41~m\,s$^{-1}$ compared with 2.71~m\,s$^{-1}$ for ERA5, together with a lower nRMSE of 44.67\% versus 50.19\%. CARRA1 also exhibits a substantially smaller negative bias, with a bias of $-0.28$~m\,s$^{-1}$ and an nBias of $-3.78\%$, compared with $-0.89$~m\,s$^{-1}$ and $-16.47\%$ for ERA5. These results indicate that CARRA1 provides a closer representation of the observed near-surface wind conditions at the selected stations, consistent with previous findings that higher-resolution regional reanalysis can reduce representativeness errors relative to ERA5 in complex terrain~\citep{copernicus_carra, yang2025danrakilometerscaledanishregional}.
\begin{table}[ht]
\centering
\caption{Station-level ACC against station observations.}
\label{tab:station_acc}
\scriptsize
\setlength{\tabcolsep}{3pt}
\begin{tabular}{llccc}
\toprule
Station ID & Station & FCN3 & HRES & GraphCast \\
\midrule
SN88690 & Hekkingen Fyr      & 0.436 & \textbf{0.474} & 0.426 \\
SN90490 & Tromsø-Langnes     & 0.377 & 0.338 & \textbf{0.382} \\
SN90760 & Fakken             & \textbf{0.375} & 0.363 & 0.318 \\
SN90800 & Torsvåg Fyr        & 0.656 & 0.635 & \textbf{0.713} \\
SN90720 & Måsvik             & 0.676 & \textbf{0.705} & 0.663 \\
SN91740 & Sørkjosen Lufthavn & 0.295 & \textbf{0.342} & 0.305 \\
SN89350 & Bardufoss          & 0.501 & 0.531 & \textbf{0.558} \\
SN87110 & Andøya             & 0.724 & 0.765 & \textbf{0.777} \\
SN85380 & Skrova Fyr         & \textbf{0.651} & 0.650 & 0.644 \\
SN94500 & Fruholmen Fyr      & 0.759 & \textbf{0.794} & 0.782 \\
SN96400 & Slettnes Fyr       & 0.720 & \textbf{0.740} & 0.694 \\
SN95350 & Banak              & 0.435 & 0.447 & \textbf{0.519} \\
\bottomrule
\end{tabular}
\smallskip
\parbox{\columnwidth}{\scriptsize
\textbf{*Bold} indicates the highest ACC at each station. Values are averaged equally across 12 forecast lead times (+6 to +72 h).}
\end{table}
\begin{table*}[tb]
\centering
\begin{threeparttable}
\caption{CARRA1 and ERA5 performance against station observations.}
\label{tab:carra1_era5}
\begin{tabular}{lccccc}
\toprule
Model & $N$ & RMSE $\downarrow$ & nRMSE $\downarrow$ & Bias $(\to 0)$ & nBias $(\to 0)$\\
\midrule
CARRA1
& 121,955
& \textbf{2.41 (0.57)}
& \textbf{44.67 (14.20)}
& \textbf{-0.28 (0.69)}
& \textbf{-3.78 (10.46)} \\

ERA5
& 122,615
& 2.71 (0.74)
& 50.19 (16.54)
& -0.89 (1.02)
& -16.47 (18.98) \\
\bottomrule
\end{tabular}
*\textbf{Bold} indicates the best result for each reference, and values are reported as the mean with the standard deviation given in parentheses. RMSE and Bias are in \(\mathrm{m\,s^{-1}}\), while nRMSE and nBias are in \%.  
\end{threeparttable}
\end{table*}
\begin{figure*}[tb]
    \centering
    \includegraphics[width=\linewidth]{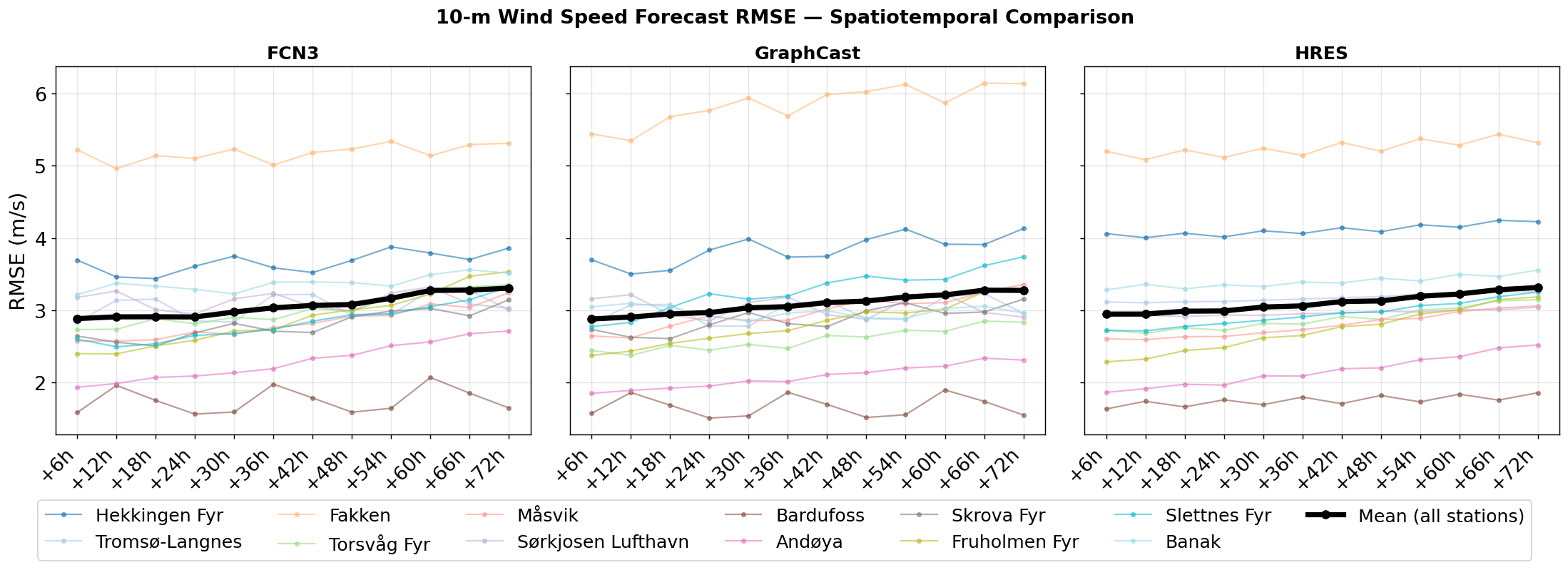}
    \caption{Station-level lead time RMSE}
    \label{fig:spatiotemporal}
\end{figure*}
\vspace{-0.5cm}
\section{Station-level ACC}
\label{app:acc_station}
Table~\ref{tab:station_acc} presents the station-level ACC against station observations. The results show considerable spatial variability, with HRES and GraphCast each achieving the highest ACC at five stations, while FCN3 performs best at two stations. Table~\ref{tab:lead_acc} presents ACC across forecast lead times. ACC generally decreases with increasing lead time for all three models. HRES performs better at shorter lead times, while GraphCast achieves higher ACC at several longer lead times.
\begin{table}[ht]
\centering
\caption{ACC by forecast lead time.}
\label{tab:lead_acc}
\small
\begin{tabular*}{\columnwidth}{@{\extracolsep{\fill}}cccc@{}}
\toprule
Lead time & FCN3 & HRES & GraphCast \\
\midrule
+6 h  & 0.598 & \textbf{0.609} & 0.599 \\
+12 h & 0.589 & \textbf{0.602} & 0.598 \\
+18 h & 0.589 & \textbf{0.595} & 0.591 \\
+24 h & 0.607 & \textbf{0.613} & 0.607 \\
+30 h & 0.574 & \textbf{0.582} & 0.572 \\
+36 h & 0.558 & \textbf{0.572} & 0.571 \\
+42 h & 0.548 & 0.555 & \textbf{0.559} \\
+48 h & 0.561 & 0.571 & \textbf{0.572} \\
+54 h & 0.517 & \textbf{0.540} & 0.536 \\
+60 h & 0.486 & 0.523 & \textbf{0.529} \\
+66 h & 0.485 & 0.505 & \textbf{0.514} \\
+72 h & 0.494 & 0.517 & \textbf{0.533} \\
\bottomrule
\end{tabular*}

\smallskip
\parbox{\columnwidth}{\footnotesize
\textbf{*Bold} indicates the highest ACC at each lead time.
Values are averaged equally across 12 stations.
}
\end{table}
\begin{table}[htbp]
\centering
\caption{Comparison of RMSE between training and post-training periods using different bootstrap block lengths.}
\label{tab:block_sensitivity}
\small
\setlength{\tabcolsep}{4pt}
\begin{tabular}{@{}lllcc@{}}
\toprule
\textbf{Model} & \textbf{$\Delta$RMSE} &
\textbf{Block} & \textbf{95\% CI} & \textbf{$p$} \\
\midrule
\multirow{4}{*}{FCN3}
& +0.010 & 1 day   & [-0.037, 0.059] & 0.651 \\
& +0.010 & 3 days  & [-0.058, 0.079] & 0.777 \\
& +0.010 & 7 days  & [-0.077, 0.096] & 0.801 \\
& +0.010 & 14 days & [-0.094, 0.118] & 0.852 \\
\midrule
\multirow{4}{*}{GC}
& -0.049 & 1 day   & [-0.100, 0.003] & 0.065 \\
& -0.049 & 3 days  & [-0.121, 0.023] & 0.185 \\
& -0.049 & 7 days  & [-0.144, 0.047] & 0.313 \\
& -0.049 & 14 days & [-0.168, 0.068] & 0.423 \\
\bottomrule
\addlinespace[3pt]
\multicolumn{5}{@{}l}{\footnotesize GC: GraphCast.} \\
\end{tabular}
\end{table}
\section{Station-level Heatmap}\label{sec:station-heatmap}
\begin{figure*}[tb]
    \centering
    \includegraphics[width=\linewidth]{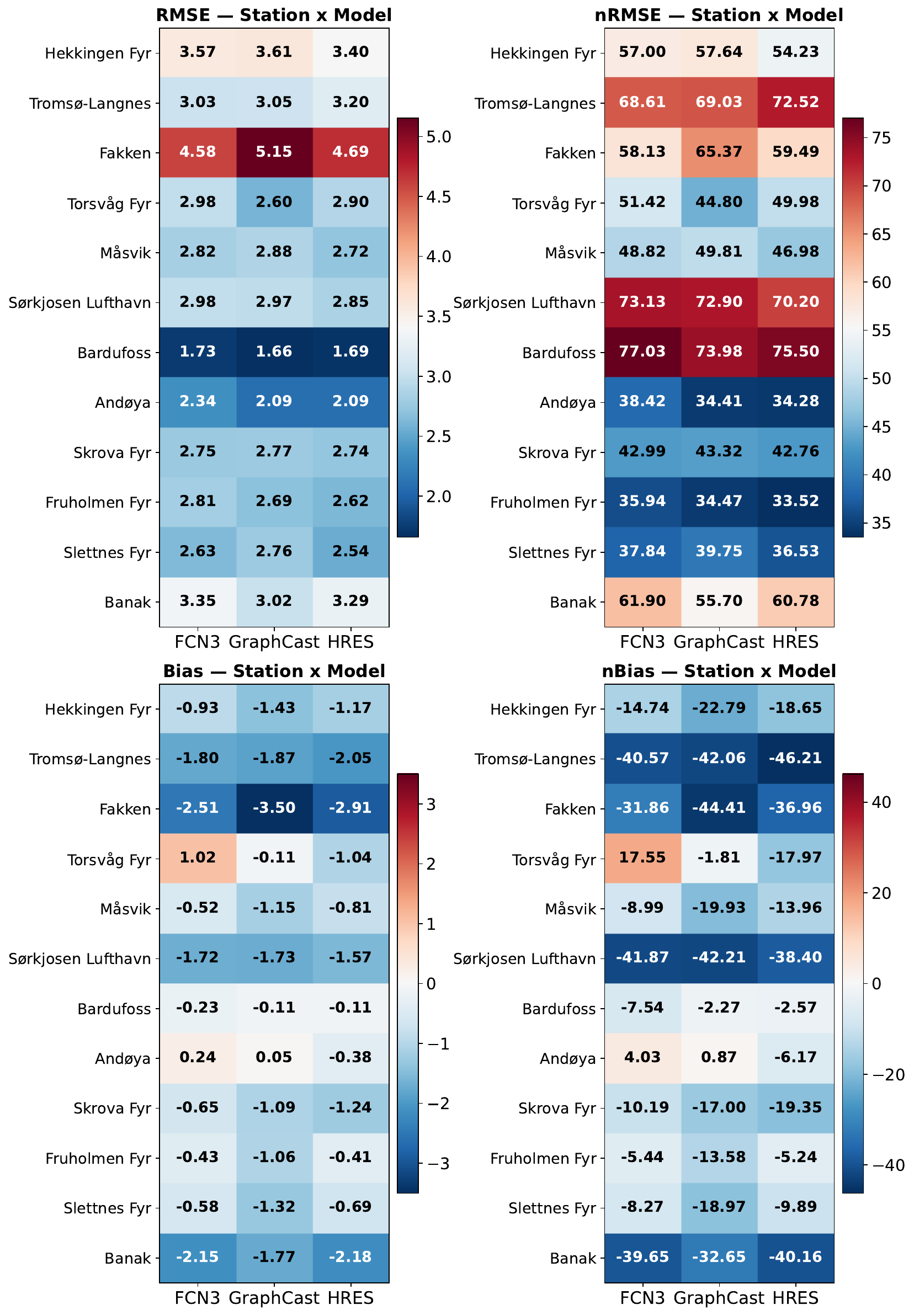}
    \caption{Station-Wise RMSE / nRMSE / Bias / nBias against Station Observations}
    \label{fig:staion-heatmap}
\end{figure*}

To complement the station-based RMSE analysis presented in the main text, we present a heatmap in Fig.~\ref{fig:staion-heatmap}, which includes RMSE, nRMSE, Bias, and nBias for FCN3, GraphCast, and HRES across the 12 stations. The nRMSE values range from 33.52\% to 77.03\% across the stations and models. The relatively high nRMSE values occur at Bardufoss, Sørkjosen Lufthavn, and Tromsø-Langnes, whereas the lower values are found at Fruholmen Fyr, Andøya, and Slettnes Fyr. Bias values range from $-3.50$ to $1.02$~m\,s$^{-1}$ and are negative for most station-model combinations. The most negative values occur at Fakken, Banak, and Tromsø-Langnes. The nBias values range from $-46.21\%$ to $17.55\%$. An interesting feature is that Bardufoss has the lowest absolute RMSE but among the highest nRMSE values. The high nRMSE partly reflects the generally low observed wind speeds at this sheltered inland station, since even relatively small absolute errors become large after normalization. In addition, the lead-time results illustrated in Fig~\ref{fig:spatiotemporal} show a diurnal variation at Bardufoss. Enhanced daytime air mixing can transport momentum from stronger winds aloft toward the surface, increasing near-surface wind speeds~\citep{lehner2019method}. Such locally driven diurnal variability may not be fully represented in the gridded fields used by the MLWP forecast models.

\section{Bootstrap Analysis of Training vs. Post-Training Performance}
\label{app:block_sensitivity}
\noindent Table~\ref{tab:block_sensitivity} presents the RMSE differences between training and post-training periods, with 95\% confidence intervals and p-values estimated using different bootstrap block lengths.

\section{Statistical Significance in High-Wind Conditions}
\label{app:highwind_significance}

The statistical significance of pairwise RMSE differences under high-wind conditions was first assessed using a station-level cluster bootstrap with 10,000 resamples. As shown in Table~\ref{tab:highwind_station_bootstrap},
only the difference between FCN3 and HRES is statistically significant.

Following the same moving block bootstrap approach described in Appendix~\ref{app:block_bootstrap}, the results are presented in Table~\ref{tab:highwind_block_bootstrap}. FCN3 has
significantly lower RMSE than both GraphCast and HRES across all block lengths, while the difference between GraphCast and HRES remains non-significant.
\begin{table}[ht]
\centering
\caption{Pairwise RMSE differences under high-wind conditions
using 10,000 bootstrap resamples.}
\label{tab:highwind_station_bootstrap}
\small
\setlength{\tabcolsep}{4pt}
\begin{tabular*}{\columnwidth}{@{\extracolsep{\fill}}lccc@{}}
\toprule
Comparison & $\Delta$RMSE & 95\% CI & Sig. \\
\midrule
FCN3--GC   & -0.2436 & [-0.507, 0.034]  & No \\
FCN3--HRES & -0.2851 & [-0.665, -0.001] & Yes \\
GC--HRES   & -0.0415 & [-0.444, 0.337]  & No \\
\bottomrule
\end{tabular*}
\parbox{\columnwidth}{\raggedright\footnotesize
GC: GraphCast.
}
\end{table}
\begin{table}[ht]
\centering
\caption{Moving block bootstrap results for pairwise RMSE
differences under high-wind conditions.}
\label{tab:highwind_block_bootstrap}
\small
\begin{threeparttable}
\begin{tabular}{lccc}
\toprule
Comparison & $\Delta$RMSE & Block & 95\% CI \\
\midrule
\multirow{4}{*}{FCN3-GC}
 & \multirow{4}{*}{-0.2436} & 1 day   & [-0.282, -0.204] \\
 &                         & 3 days  & [-0.285, -0.200] \\
 &                         & 7 days  & [-0.286, -0.198] \\
 &                         & 14 days & [-0.287, -0.197] \\
\midrule
\multirow{4}{*}{FCN3-HRES}
 & \multirow{4}{*}{-0.2851} & 1 day   & [-0.334, -0.231] \\
 &                         & 3 days  & [-0.339, -0.225] \\
 &                         & 7 days  & [-0.341, -0.219] \\
 &                         & 14 days & [-0.342, -0.218] \\
\midrule
\multirow{4}{*}{GC-HRES}
 & \multirow{4}{*}{-0.0415} & 1 day   & [-0.089, 0.007] \\
 &                         & 3 days  & [-0.093, 0.015] \\
 &                         & 7 days  & [-0.097, 0.018] \\
 &                         & 14 days & [-0.094, 0.016] \\
\bottomrule
\end{tabular}
\begin{tablenotes}[flushleft]
\footnotesize
\item GC: GraphCast. Negative $\Delta$RMSE indicates
lower RMSE for the first model in each comparison.
\end{tablenotes}
\end{threeparttable}
\end{table}
\section{Station-Level Scatter Plots}\label{sec:scatter-plots}
\begin{figure}[tbhp]
    \centering
    \includegraphics[width=\linewidth]{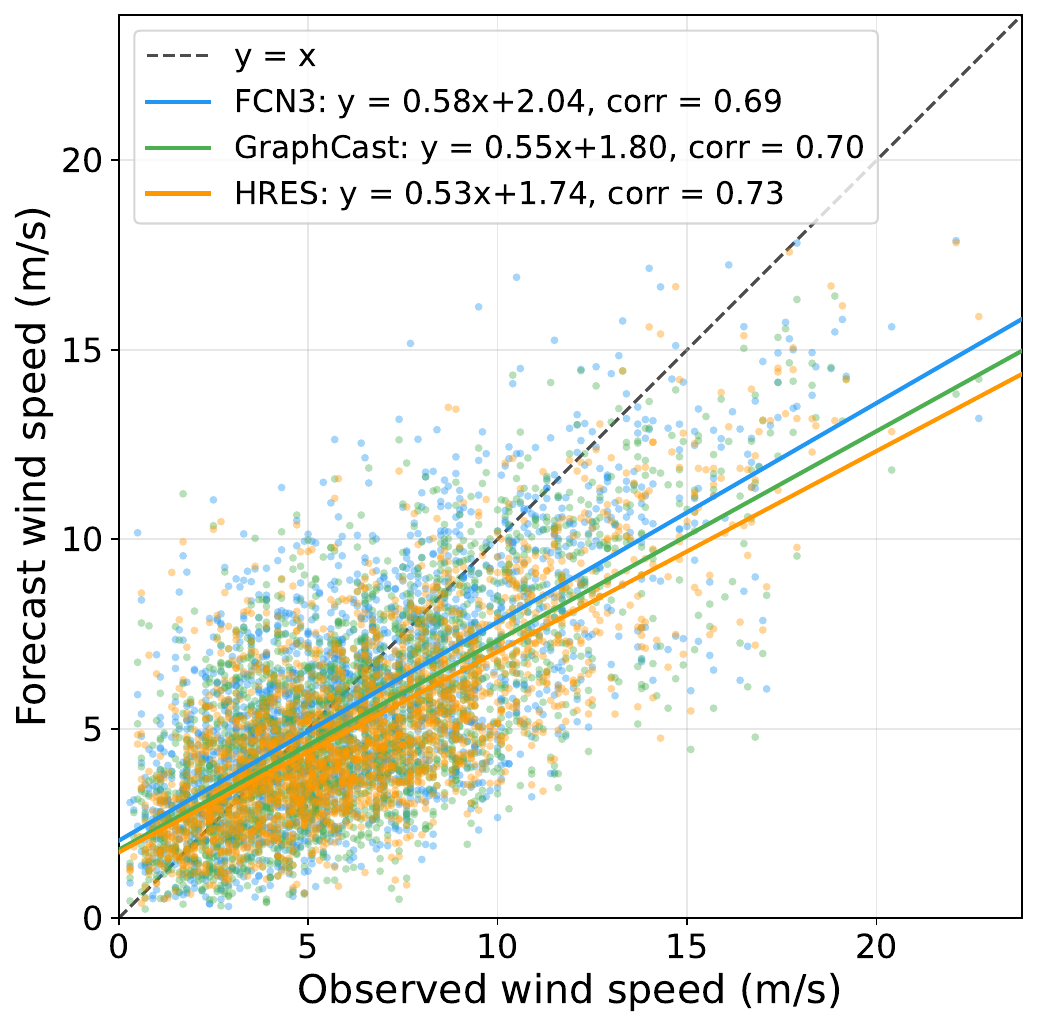}
    \caption{Scatter plot for Skrova Fyr.}
    \label{fig:hw_SN85380}
\end{figure}

\begin{figure}[tbhp]
    \centering
    \includegraphics[width=\linewidth]{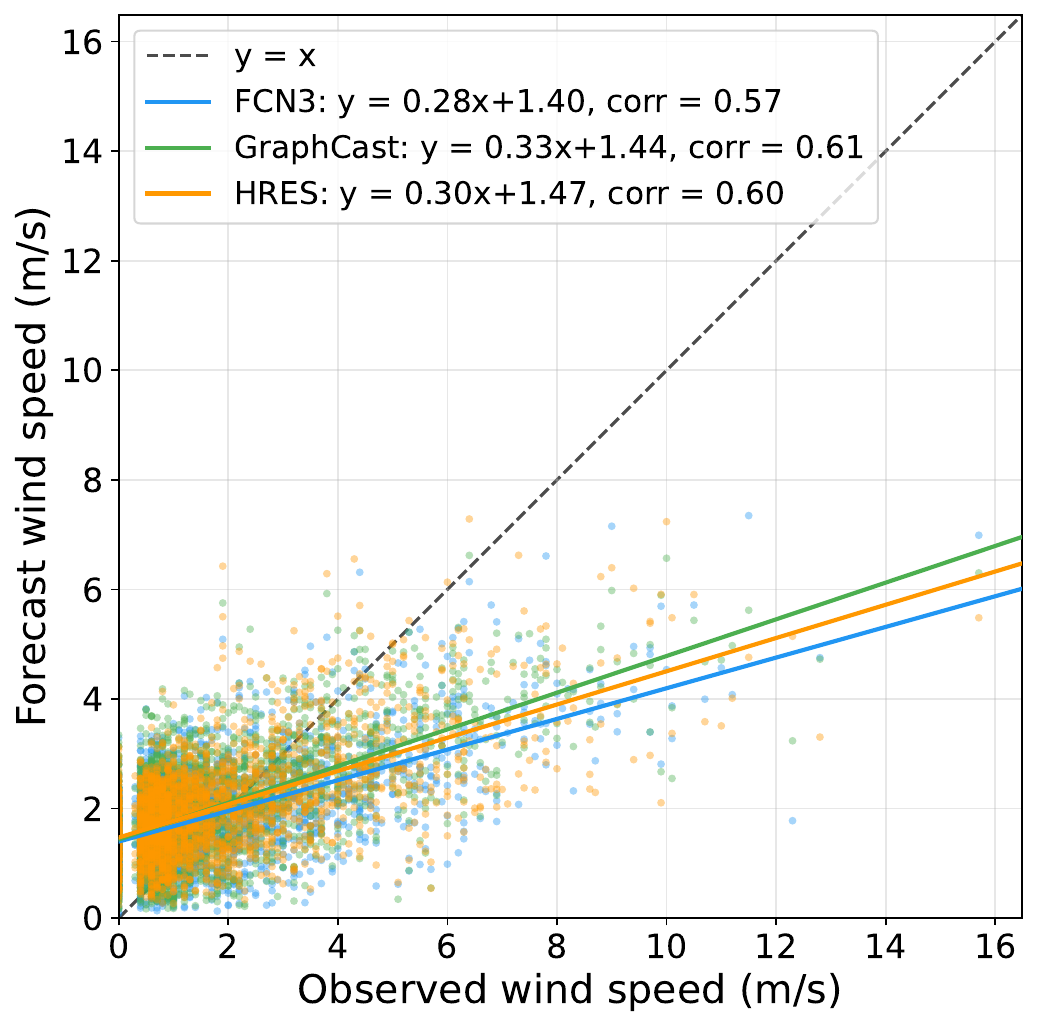}
    \caption{Scatter plot for Bardufoss.}
    \label{fig:hw_SN89350}
\end{figure}

\begin{figure*}[tbhp]
    \centering

    \begin{subfigure}[t]{0.47\textwidth}
        \centering
        \includegraphics[width=\linewidth]{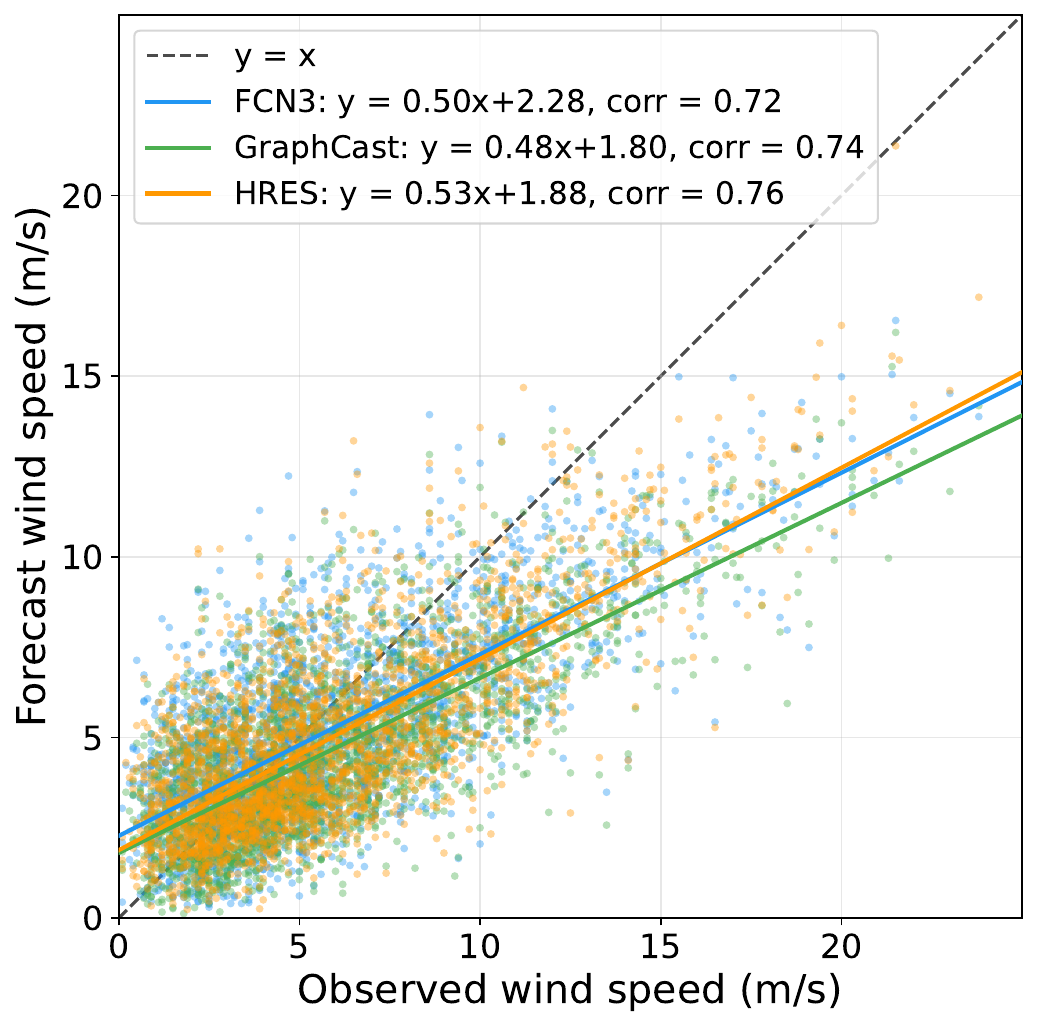}
        \caption{Måsvik (SN90720)}
        \label{fig:hw_SN90720}
    \end{subfigure}
    \hfill
    \begin{subfigure}[t]{0.47\textwidth}
        \centering
        \includegraphics[width=\linewidth]{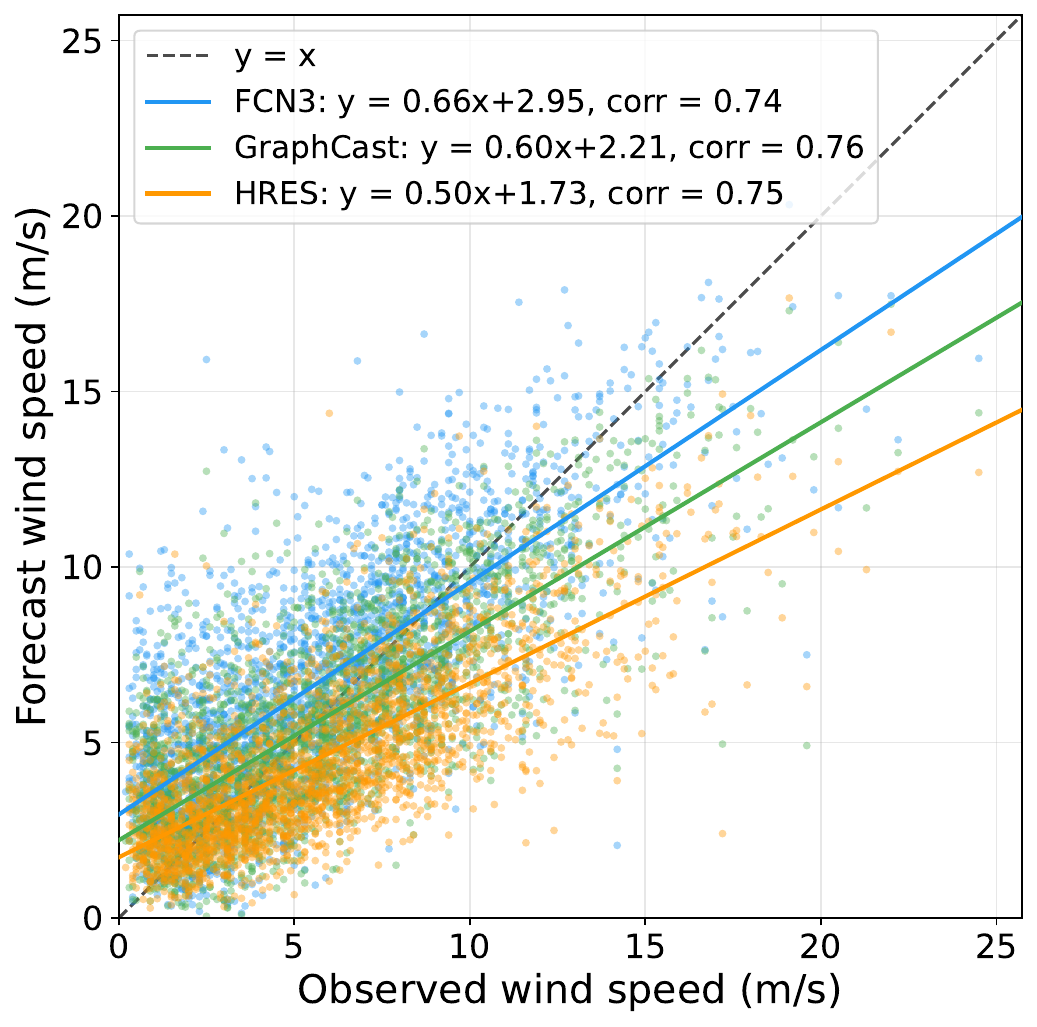}
        \caption{Torsvåg Fyr (SN90800)}
        \label{fig:hw_SN90800}
    \end{subfigure}

    \begin{subfigure}[t]{0.49\textwidth}
        \centering
        \includegraphics[width=\linewidth]{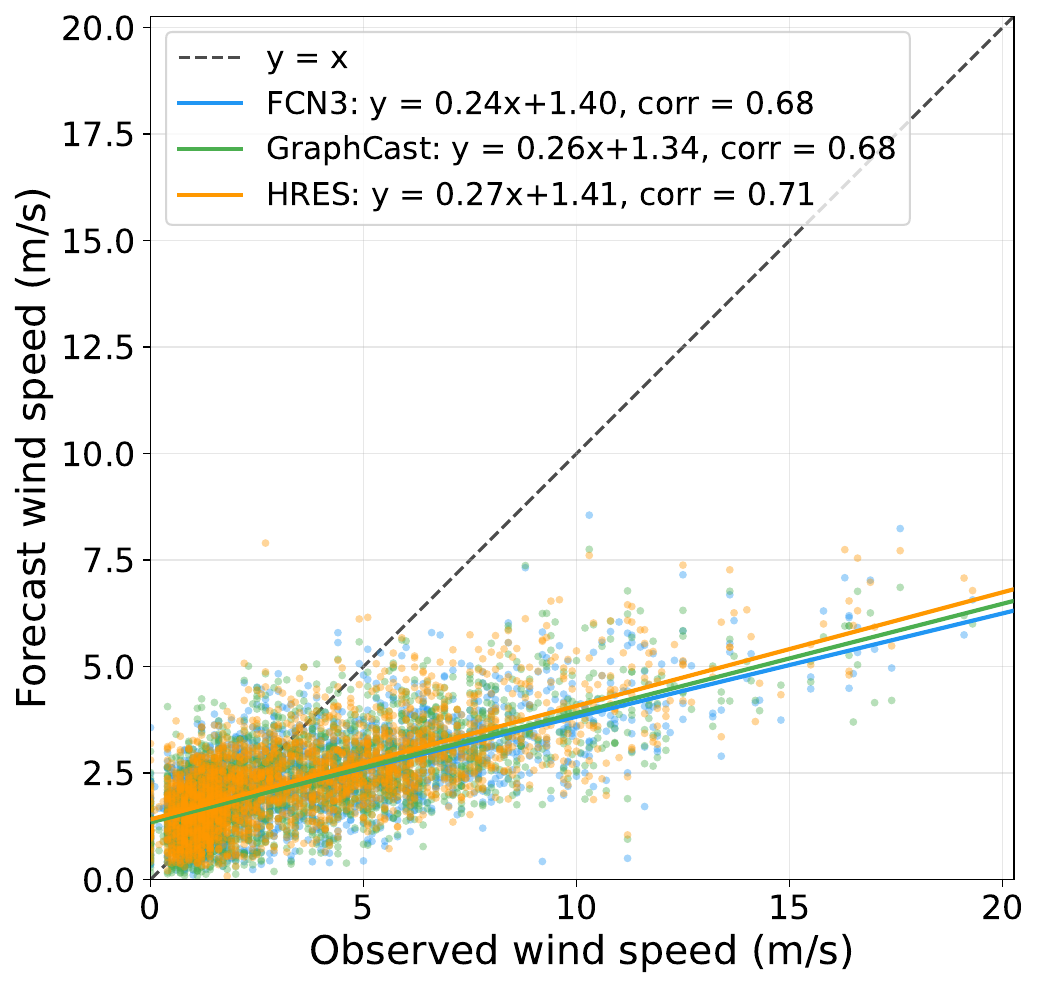}
        \caption{Sørkjosen Lufthavn (SN91740)}
        \label{fig:hw_SN91740}
    \end{subfigure}
    \hfill
    \begin{subfigure}[t]{0.47\textwidth}
        \centering
        \includegraphics[width=\linewidth]{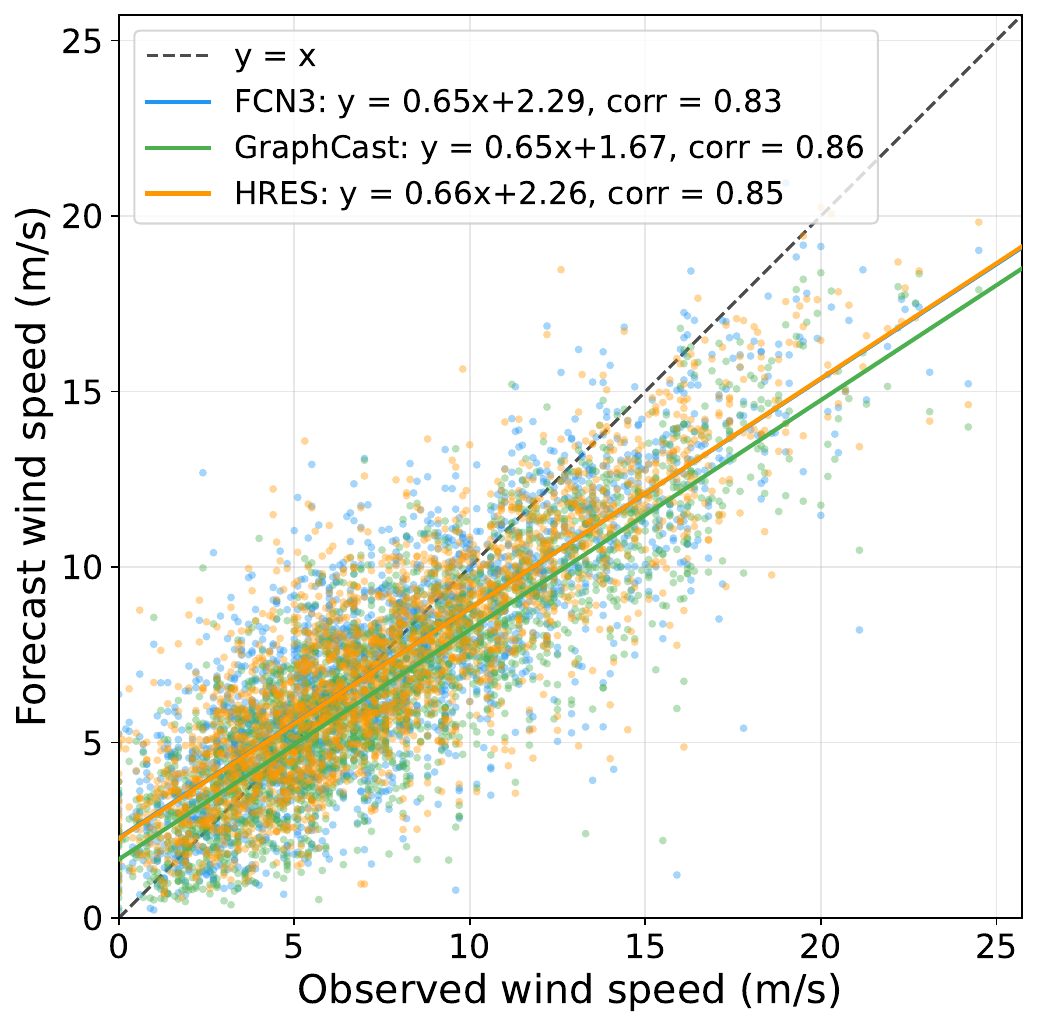}
        \caption{Fruholmen Fyr (SN94500)}
        \label{fig:hw_SN94500}
    \end{subfigure}

    \begin{subfigure}[t]{0.47\textwidth}
        \centering
        \includegraphics[width=\linewidth]{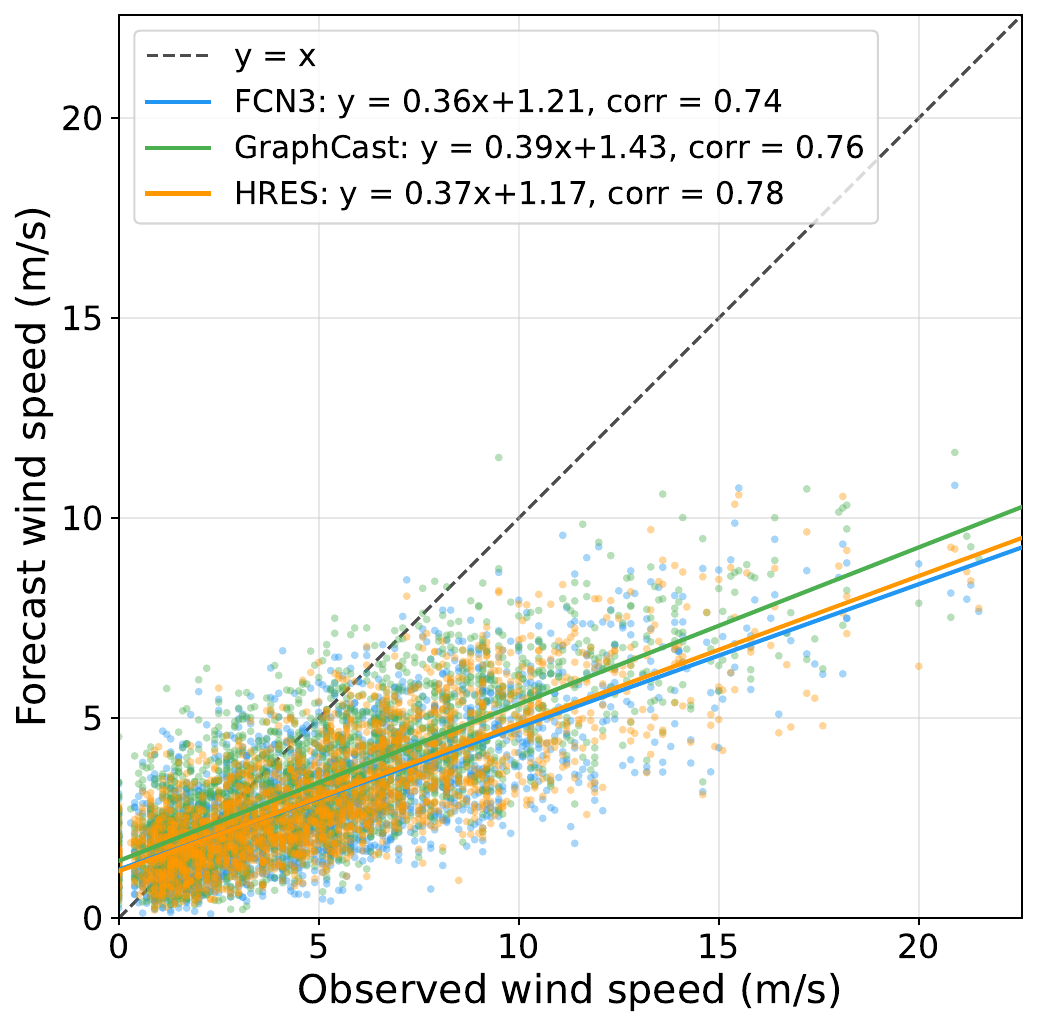}
        \caption{Banak (SN95350)}
        \label{fig:hw_SN95350}
    \end{subfigure}
    \hfill
    \begin{subfigure}[t]{0.47\textwidth}
        \centering
        \includegraphics[width=\linewidth]{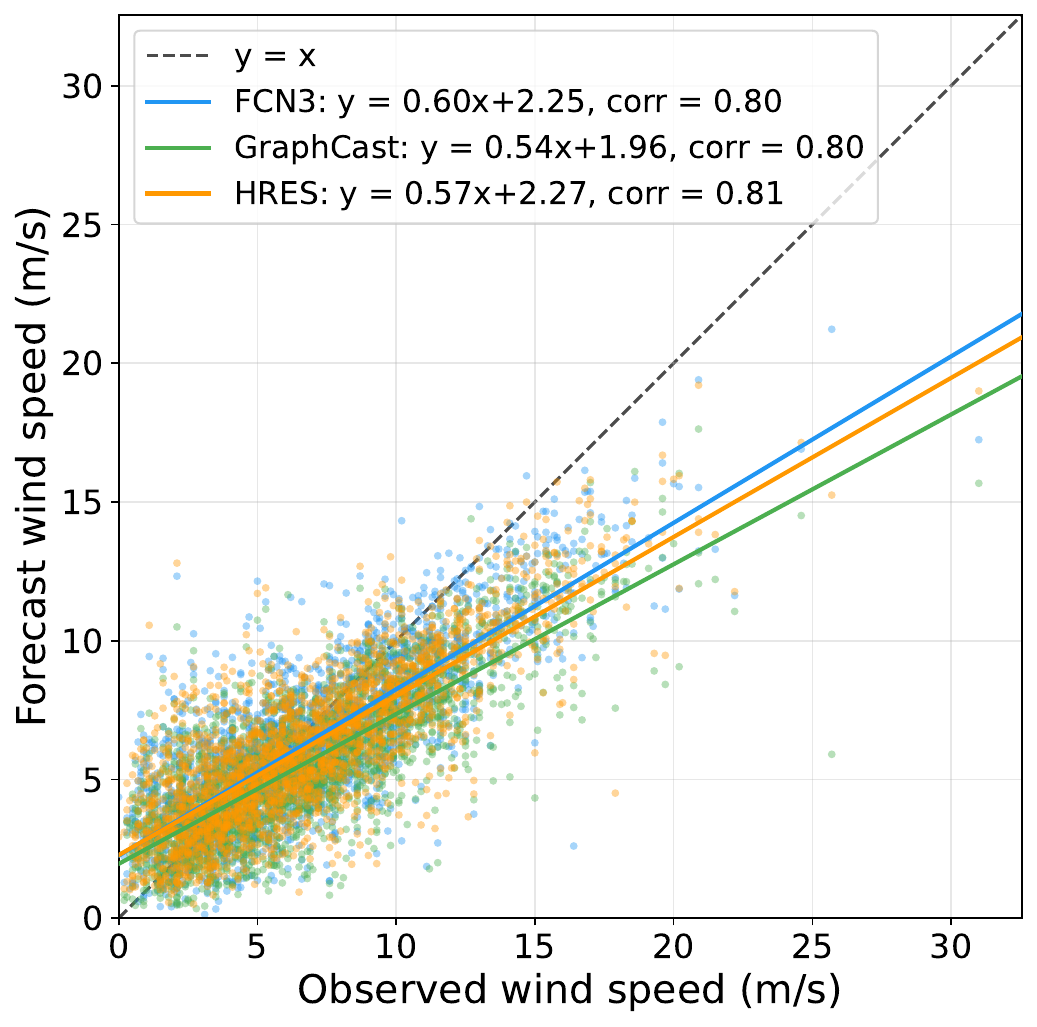}
        \caption{Slettnes Fyr (SN96400)}
        \label{fig:hw_SN96400}
    \end{subfigure}

    \caption{Scatter plots of observed and predicted wind speeds at stations.}
    \label{fig:hw_scatter_appendix}
\end{figure*}

We present the scatter plots for the remaining stations as mentioned ((Figs.~\ref{fig:hw_SN85380}, \ref{fig:hw_SN89350}, and \ref{fig:hw_scatter_appendix})), comparing observed wind speeds with predictions from FCN3, GraphCast, and HRES. These plots provide a station-specific view of model performance.

\begin{table*}[t]
\centering
\caption{Station-level wind forecast performance of FCN3, GraphCast (GC), and HRES across wind-speed categories defined by the Beaufort scale.}
\label{tab:beaufort_detailed}
\fontsize{7.5}{9}\selectfont
\setlength{\tabcolsep}{2.5pt}
\renewcommand{\arraystretch}{1.06}

\begin{tabular*}{\textwidth}{@{\extracolsep{\fill}}clccccccc@{}}
\toprule
& & & \multicolumn{3}{c}{RMSE $\downarrow$} & \multicolumn{3}{c}{Bias $(\to 0)$} \\
\cmidrule(lr){4-6}\cmidrule(l){7-9}
\textbf{Beaufort} & \textbf{Range} & \textbf{$N_{\mathrm{obs}}$}
& \textbf{FCN3} & \textbf{GC} & \textbf{HRES}
& \textbf{FCN3} & \textbf{GC} & \textbf{HRES} \\
\midrule
0  & $<0.5$     & 7,166   & 2.928 (1.171) & \textbf{2.443 (0.857)} & 2.612 (0.947) & +2.593 (1.027) & \textbf{+2.150 (0.763)} & +2.352 (0.849) \\
1  & 0.5--1.5   & 39,338  & 2.507 (1.091) & \textbf{2.082 (0.735)} & 2.144 (0.872) & +1.931 (0.944) & \textbf{+1.521 (0.622)} & +1.644 (0.749) \\
2  & 1.6--3.3   & 68,116  & 2.046 (0.785) & \textbf{1.686 (0.503)} & 1.704 (0.578) & +0.889 (0.942) & \textbf{+0.497 (0.639)} & +0.605 (0.741) \\
3  & 3.4--5.4   & 81,042  & 2.162 (0.233) & 1.943 (0.207) & \textbf{1.926 (0.226)} & \textbf{-0.298 (1.168)} & -0.674 (0.855) & -0.622 (0.937) \\
4  & 5.5--7.9   & 78,381  & 2.752 (0.634) & 2.720 (0.666) & \textbf{2.697 (0.716)} & \textbf{-1.544 (1.486)} & -1.919 (1.182) & -1.930 (1.205) \\
5  & 8.0--10.7  & 52,295  & \textbf{3.719 (1.244)} & 3.821 (1.190) & 3.844 (1.233) & \textbf{-2.853 (1.874)} & -3.241 (1.573) & -3.313 (1.316) \\
6  & 10.8--13.8 & 25,655  & \textbf{4.963 (1.843)} & 5.162 (1.728) & 5.215 (1.768) & \textbf{-4.280 (2.318)} & -4.699 (2.026) & -4.790 (2.006) \\
7  & 13.9--17.1 & 10,212  & \textbf{6.344 (2.410)} & 6.673 (2.297) & 6.723 (2.317) & \textbf{-5.756 (2.787)} & -6.277 (2.515) & -6.337 (2.512) \\
8  & 17.2--20.7 & 3,249   & \textbf{7.653 (2.980)} & 8.131 (2.777) & 8.048 (2.563) & \textbf{-7.149 (3.230)} & -7.789 (2.908) & -7.644 (2.728) \\
9  & 20.8--24.4 & 873     & \textbf{9.206 (2.778)} & 9.873 (2.768) & 9.522 (2.782) & \textbf{-8.738 (2.779)} & -9.629 (2.755) & -9.161 (2.885) \\
10 & 24.5--28.4 & 209     & \textbf{13.171 (2.918)} & 14.284 (3.089) & 13.639 (2.813) & \textbf{-12.911 (3.103)} & -14.139 (3.139) & -13.426 (2.934) \\
11 & 28.5--32.6 & 46      & 17.556 & 18.153 & \textbf{17.474} & -17.412 & -18.085 & \textbf{-17.332} \\
12 & $\geq 32.7$ & 3       & --- & --- & --- & --- & --- & --- \\
\midrule
\textbf{All} & All & 366,585 & 2.957 (0.707) & 2.938 (0.861) & \textbf{2.893 (0.754)} & \textbf{-0.852 (1.039)} & -1.258 (0.988) & -1.214 (0.851) \\
\bottomrule
\end{tabular*}

\vspace{6pt}

\begin{tabular*}{\textwidth}{@{\extracolsep{\fill}}ccccccc@{}}
\toprule
& \multicolumn{3}{c}{nRMSE $\downarrow$} & \multicolumn{3}{c}{nBias $(\to 0)$} \\
\cmidrule(lr){2-4}\cmidrule(l){5-7}
\textbf{Beaufort} & \textbf{FCN3} & \textbf{GC} & \textbf{HRES}
& \textbf{FCN3} & \textbf{GC} & \textbf{HRES} \\
\midrule
0  & 1558.0 (971.4) & \textbf{1358.2 (871.6)} & 1426.2 (916.2) & +1376.0 (832.9) & \textbf{+1189.4 (743.9)} & +1288.7 (836.2) \\
1  & 228.6 (99.4) & \textbf{190.1 (66.8)} & 195.7 (79.4) & +176.1 (86.5) & \textbf{+139.0 (57.3)} & +150.2 (68.7) \\
2  & 82.0 (31.0) & \textbf{67.7 (19.7)} & 68.3 (22.2) & +35.1 (38.2) & \textbf{+19.6 (26.0)} & +23.8 (29.8) \\
3  & 49.3 (5.2) & 44.3 (4.6) & \textbf{43.9 (5.1)} & \textbf{-6.9 (26.8)} & -15.4 (19.6) & -14.3 (21.5) \\
4  & 41.7 (9.9) & 41.2 (10.3) & \textbf{40.9 (11.1)} & \textbf{-23.5 (22.7)} & -29.2 (18.1) & -29.3 (18.5) \\
5  & \textbf{40.7 (13.9)} & 41.8 (13.2) & 42.1 (13.8) & \textbf{-31.3 (20.7)} & -35.5 (17.4) & -36.3 (17.0) \\
6  & \textbf{41.5 (15.6)} & 43.1 (14.6) & 43.6 (15.0) & \textbf{-35.8 (19.5)} & -39.2 (17.1) & -40.0 (17.0) \\
7  & \textbf{42.0 (15.9)} & 44.1 (15.1) & 44.5 (15.4) & \textbf{-38.1 (18.4)} & -41.5 (16.7) & -41.9 (16.7) \\
8  & \textbf{41.3 (15.9)} & 43.8 (14.7) & 43.4 (13.7) & \textbf{-38.5 (17.3)} & -42.0 (15.4) & -41.3 (14.6) \\
9  & \textbf{41.5 (12.5)} & 44.5 (12.4) & 43.0 (12.6) & \textbf{-39.4 (12.6)} & -43.4 (12.3) & -41.3 (13.0) \\
10 & \textbf{51.1 (11.3)} & 55.4 (12.0) & 52.9 (11.1) & \textbf{-50.1 (12.1)} & -54.9 (12.1) & -52.1 (11.5) \\
11 & 58.7 & 60.7 & 58.4 & -58.2 & -60.4 & \textbf{-58.0} \\
12 & --- & --- & --- & --- & --- & --- \\
\midrule
\textbf{All} & 54.1 (14.2) & 53.4 (14.6) & \textbf{53.1 (15.0)} & \textbf{-15.6 (19.3)} & -21.4 (16.7) & -21.3 (15.8) \\
\bottomrule
\end{tabular*}

\vspace{4pt}
\parbox{\textwidth}{\raggedright\fontsize{7}{8.5}\selectfont
Events are classified by observed wind speed. $N_{\mathrm{obs}}$ is the number of observations. Metrics are computed per station and forecast lead time, then averaged. Values in parentheses denote the standard deviation across stations, calculated at each lead time and averaged over 12 leads. nRMSE and nBias are normalized by the mean observed wind speed in each category. Bias is forecast minus observation. Bold indicates the smallest RMSE/nRMSE or absolute bias/nBias in each row. Dashes indicate unavailable estimates.
}
\end{table*}

\end{document}